\documentclass[pmlr]{jmlr}

\usepackage{longtable}

 \usepackage{booktabs}
 
\usepackage[load-configurations=version-1]{siunitx} 

\makeatletter
\def\set@curr@file#1{\def\@curr@file{#1}} 
\makeatother

\usepackage{paralist}
\usepackage{amssymb}
\usepackage{graphicx}
\usepackage{mathtools}
\usepackage{bbm}
\usepackage{xcolor}
\usepackage{hyperref}
\usepackage{algorithmic}
\usepackage{caption}
\usepackage{enumitem}
\usepackage{amsmath}

\usepackage{tikz}
\usetikzlibrary{shapes,arrows,fit,chains,matrix}
\usetikzlibrary{positioning, shapes.geometric}
\usetikzlibrary{bayesnet}
\tikzstyle{node_no_draw} = [circle, minimum size = 10mm]
\tikzstyle{node} = [circle, minimum size = 10mm, draw =black!80]
\tikzstyle{nodeinter} = [rectangle, minimum size = 10mm, draw =black!80, fill=gray!30]
\tikzstyle{nodeinterwhite} = [rectangle, minimum size = 10mm, draw =black!80]
\tikzstyle{nodeobserved} = [circle, minimum size = 10mm, draw =black!80, fill=gray!30]
\tikzstyle{box} = [rectangle, draw =black!0]
\tikzstyle{arrow} = [thick,->,>=stealth,line width=0.3mm]
\tikzstyle{arrow2} = [dashed,->,>=stealth,line width=0.6mm]

\usetikzlibrary{calc}
\makeatletter
\tikzset{
  prefix after node/.style={
    prefix after command={\pgfextra{#1}}
  },
  /semifill/ang/.store in=\semi@ang,
  /semifill/ang=0,
  semifill/.style={
    circle, draw,
    prefix after node={
      \typeout{aaa \semi@ang}
      \let\nodename\tikz@last@fig@name
      \fill[/semifill/.cd, /semifill/.search also={/tikz}, #1]
        let \p1 = (\nodename.north), \p2 = (\nodename.center) in
        let \n1 = {\y1 - \y2} in
        (\nodename.\semi@ang) arc [radius=\n1, start angle=\semi@ang, delta angle=180];
    },
  }
}
\makeatother

\newcommand{\given}{{\, | \,}}
\newcommand{\EE}{\mathbb{E}}

\newcommand{\xv}{{\mathbf X}}
\newcommand{\yv}{{\mathbf Y}}

\newcommand{\yvs}{Y}
\newcommand{\zv}{\mathbf{Z}}
\newcommand{\sz}{\boldsymbol{z}}

\newcommand{\thetav}{\boldsymbol{\theta}}
\newcommand{\omegav}{\boldsymbol{\omega}}

\newcommand{\psiv}{\boldsymbol{\psi}}
\newcommand{\epsilonv}{\boldsymbol{\epsilon}}

\newcommand{\phiv}{\boldsymbol{\phi}}

\def \d {K}
\def \l {k}

\def \Y {\mathbf{Y}}

\def \y {\boldsymbol{y}}

\newcommand{\defeq}{\vcentcolon=}

\newcolumntype{P}[1]{>{\centering\arraybackslash}p{#1}}

\theorembodyfont{\upshape}
\theoremheaderfont{\scshape}
\theorempostheader{:}
\theoremsep{\newline}

\usepackage{wrapfig}

\jmlrvolume{}
\jmlryear{2023}
\jmlrworkshop{Machine Learning for Healthcare}

\title[AIRIVA: A Deep Generative Model of Adaptive Immune Repertoires]{AIRIVA: A Deep Generative Model of \\ Adaptive Immune Repertoires}

\author{\Name{Melanie F. Pradier}
       \Email{melanief@microsoft.com}\\ 
       \addr Microsoft Research, Cambridge, UK\\
       \Name{Niranjani Prasad$^\star$}
       \Email{niranjani.prasad@microsoft.com}\\ 
       \addr Microsoft Research, Cambridge, UK\\
       \Name{Paidamoyo Chapfuwa$^\star$}
       \Email{pchapfuwa@microsoft.com}\\ 
       \addr Microsoft Research, Redmond, USA\\
       \Name{Sahra Ghalebikesabi}
       \Email{Sahra.Ghalebikesabi@univ.ox.ac.uk}\\ 
       \addr Oxford University, Oxford, UK\\
       \Name{Max Ilse}
       \Email{max.ilse@microsoft.com}\\ 
       \addr Microsoft Research, Amsterdam, The Netherlands\\
       \Name{Steven Woodhouse}
       \Email{swoodhouse@adaptivebiotech.com}\\ 
       \addr Adaptive Biotechnologies, Redmond, USA\\
       \Name{Rebecca Elyanow}
       \Email{relyanow@adaptivebiotech.com}\\ 
       \addr Adaptive Biotechnologies, Redmond, USA\\
       \Name{Javier Zazo}
       \Email{javierzazo@microsoft.com}\\ 
       \addr Microsoft Research, Cambridge, UK\\
       \Name{Javier Gonzalez Hernandez}
       \Email{javierzazo@microsoft.com}\\ 
       \addr Microsoft Research, Cambridge, UK\\
       \Name{Julia Greissl}
       \Email{julia.greissl@microsoft.com}\\ 
       \addr Microsoft Research, Redmond, USA\\
       \Name{Edward Meeds}
       \Email{ted.meeds@microsoft.com}\\ 
       \addr Microsoft Research, Cambridge, UK\\
       } 

\editor{Editor's name}

\begin{document}

\maketitle

\begin{abstract}
Recent advances in immunomics have shown that T-cell receptor (TCR) signatures can accurately predict active or recent infection by leveraging the high specificity of TCR binding to disease antigens.
However, the extreme diversity of the adaptive immune repertoire presents challenges in reliably identifying disease-specific TCRs. Population genetics and sequencing depth can also have strong systematic effects on repertoires, which requires careful consideration when developing diagnostic models.
We present an Adaptive Immune Repertoire-Invariant Variational Autoencoder (AIRIVA), a generative model that learns a low-dimensional, interpretable, and compositional representation of TCR repertoires to disentangle such systematic effects in repertoires.
%
We apply AIRIVA to two infectious disease case-studies: COVID-19 (natural infection and vaccination) and the Herpes Simplex Virus (HSV-1 and HSV-2), and empirically show that we can disentangle the individual disease signals. We further demonstrate AIRIVA's capability to: learn from unlabelled samples; generate in-silico TCR repertoires by intervening on the latent factors; and identify disease-associated TCRs validated using TCR annotations from external assay data.

\end{abstract}


\section{Introduction} \label{sec:introduction}
Precision medicine, that is diagnosis and treatment targeted to an individual, is one of the most promising applications of machine learning (ML) in healthcare. Because of an explosion in biomarkers and corresponding large datasets, there has been increased interest in applying deep learning artificial intelligence (AI) models in a clinical setting~\citep{miotto2018deep, sandeep2020deep}. In radiology and pathology settings, AI-based diagnostics and human-in-the-loop systems are now common~\citep{rajpurkar2022, oktay2020evaluation}. 
Data complexity and clinical risk have spurred the development of ML models that are robust, interpretable and fair. This includes works that explicitly incorporate interpretability and resilience to distribution shifts in real-world domains such as longitudinal biomarker modeling~\citep{hussain2021neural}, biomarker discovery~\citep{pradier2019case} and medical imaging~\citep{chartsias2019disentangled, ilse2020diva}.


\begin{figure}
     \centering
     \includegraphics[width=0.6\linewidth]{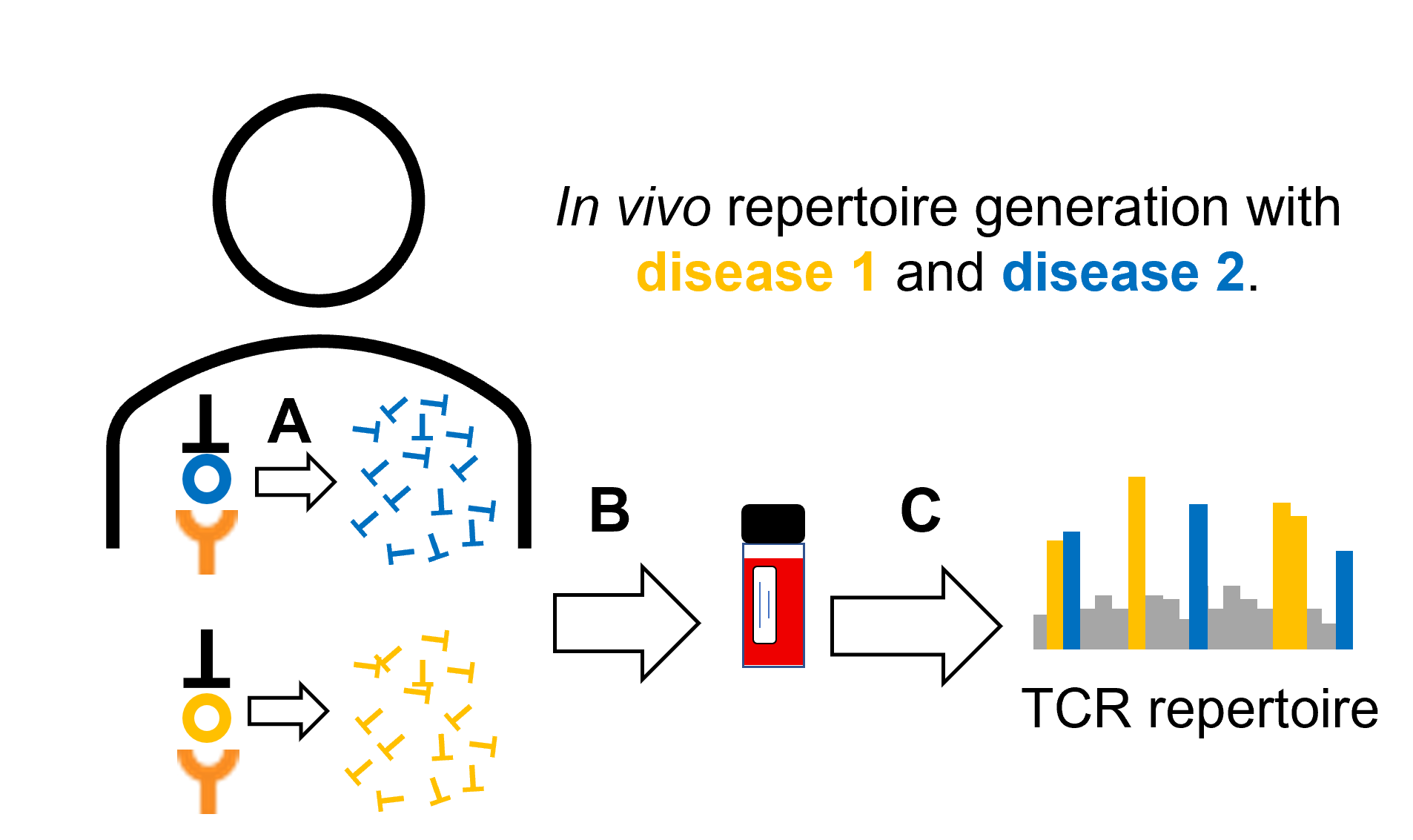}
    
      \caption{
      {\bf In-vivo generative process of immune repertoires.} \textit{Antigen}s are presented by {\em human leukocyte antigens} (HLAs) to T cells. When a T cell binds to a specific antigen-HLA complex, it undergoes clonal expansion (A), which is captured in a blood sample (B) and quantified via high-throughput immuno-sequencing resulting in a TCR \textit{repertoire} (C).
     }
     \label{fig:intro-antigen1}
     \vspace{-0.5cm}
\end{figure}

Our focus is {\em \bf immunomics}, where recent advances in high-throughput T Cell sequencing~\citep{robinsimmunoseq} have opened the door to a new form of precision diagnostic based on adaptive immune repertoires (see Figure~\ref{fig:intro-antigen1}).  Immunomics-based diagnostic models have shown promise in
detecting new or past infection in Cytomegalovirus~\citep{emerson_immunosequencing_2017}, Lyme~\citep{greissl2021immunosequencing}, and COVID-19~\citep{snyder2020magnitude}. 
With the promise of a new form of precision diagnostics comes the challenge of validating signal and
building clinical trust within an extremely complex, diverse, high-dimensional data domain that lacks human-understandable structure.  

\begin{figure*}[]
    \centering
    \includegraphics[width=1.0\linewidth]{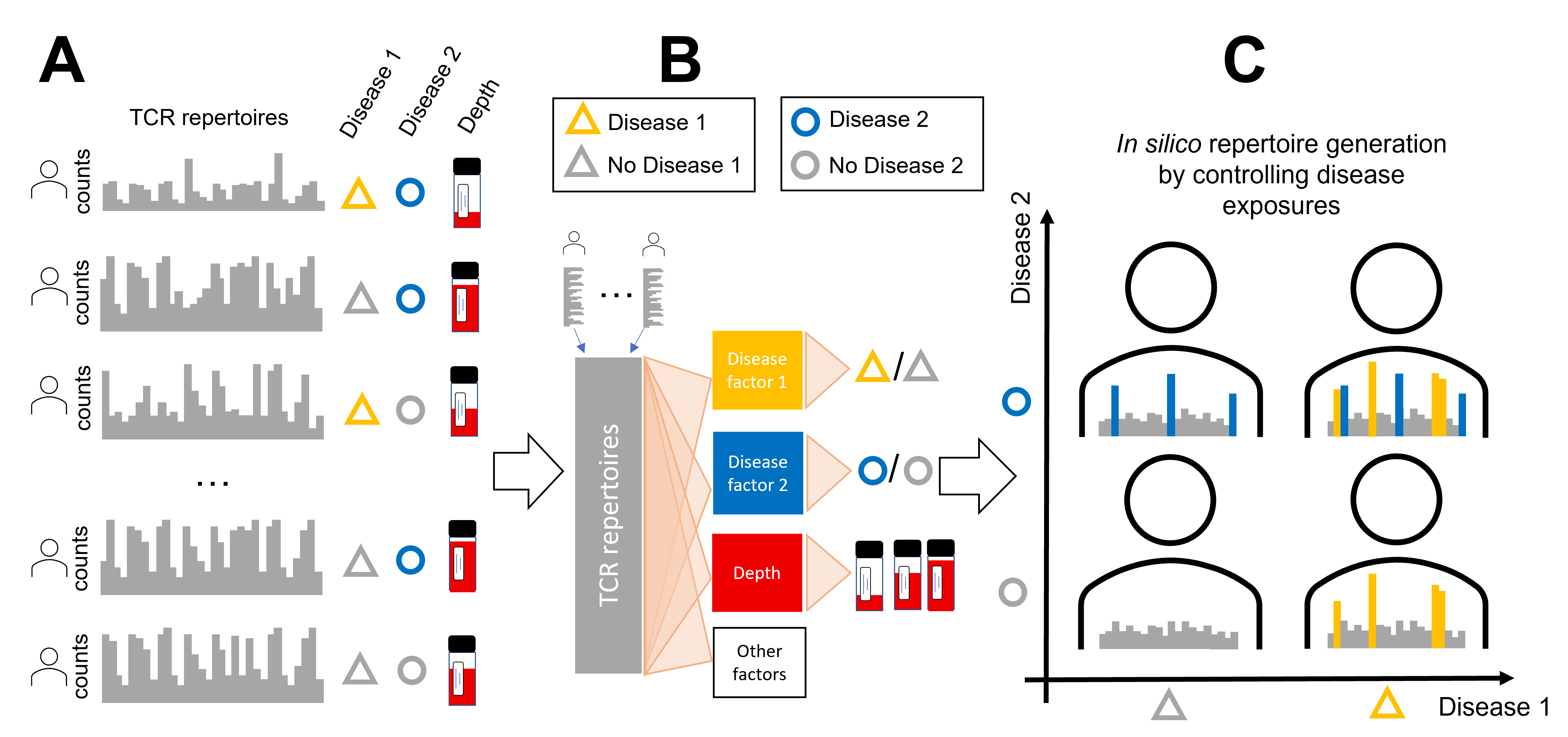}
     \caption{{\bf Predictive and generative data flow in AIRIVA.}   
(A) Training data consists of TCR counts and repertoire labels. (B) TCRs are mapped to independent latent factors (factorized representation) which are subsequently used for prediction. (C) By intervening on targets and depth (toggles of a simulator), in-silico repertoires are generated.  
    }
    \label{fig:intro-antigen2}
\end{figure*}

To address these challenges, a {\em generative} model framework that explicitly incorporates {\em interpretable} latent factors linked to repertoire {\em targets} (e.g. disease labels, genetic variants, sequencing depth, batch id) is highly desirable. We propose the {\bf Adaptive Immune Repertoire-Invariant Autoencoder} (AIRIVA) as a solution:
a deep generative model for TCR repertoires, an adaptation of recent work~\citep{ilse2020diva, joy2020capturing} to immunomics.  AIRIVA is a semi-supervised generative model trained to explicitly learn disentangled and interpretable latent representations of TCR repertoires and can be used to both predict multiple disease labels and to generate in-silico repertoires, which can in turn be used for assigning TCR-disease associations.  Figure~\ref{fig:intro-antigen2} illustrates these key ideas.

\subsection*{Generalizable Insights about Machine Learning in the Context of Healthcare}
%
\begin{itemize}
\itemsep0em
    \item Immunomics is a relatively new field that has the potential to revolutionize precision medicine. We propose a generative model for handling complex immunomics data which is high dimensional, sparse, and heterogeneous.
    \item By enforcing a disentangled latent representation, we are able to learn interpretable \emph{label-specific} factors of variation. 
    \item We propose an approach for generating in-silico (synthetic) \emph{counterfactual} repertoires with AIRIVA by \emph{intervening} on specific latent factors of interest. We leverage the generated counterfactuals for model selection and identification of label-specific TCRs, validated with an external dataset of TCR annotations.
    \item We empirically demonstrate AIRIVA’s capacity to disentangle disease signal in the context of two infectious diseases, COVID-19 and Herpes Simplex Virus, improving model robustness across subgroups of interest.
   \item We demonstrate improved AIRIVA label prediction by learning with additional TCR repertoires with missing labels. Leveraging samples with missing labels is crucial for the immunomics setting where cohorts are heterogeneous with small sample sizes.
\end{itemize}

\section{Immunomics}\label{sec:immuno}
T cells are a central component of the adaptive immune response that is complementary to antibodies secreted by B cells. Their role is to identify and destroy foreign or cancerous cells that are expressing non-native antigens --- small fragments of proteins presented by {\em human leukocyte antigens} (HLAs) --- and to recognize antigens collected from other immune cells as self or non-self,  to regulate the overall response~\citep{murphy2016janeway}. The adaptive immune system ``learns" 
to recognize antigens by first generating random, non-self binding naive cells, which then undergo clonal expansion upon encountering their cognate antigen~\citep{malissen2007}. 
Thus an immune repertoire, sampled from $O(10^7)$ unique TCRs \citep{emerson_immunosequencing_2017} in an individual, is a mixture of naive (random) and expanded (antigen-specific) T cells.  Individuals with shared HLAs who are exposed to similar antigens (e.g. viral infections, vaccinations), tend to share similar and even identical TCRs~\citep{Dendrou2018}; this mechanism underpins statistical immunomics modeling~\citep{GREIFF2020109} (see Section~\ref{sec:eslg}). 




\subsection{Modeling Challenges}
High-throughput sequencing given a standard blood sample~\citep{robinsimmunoseq} has opened the door to a new form of precision diagnostics. The complexity and diversity of repertoires present several important modeling challenges, see~\citep{pavlovic2022using} for a detailed overview.
%
Repertoire composition is influenced by many factors, primarily by antigen {\bf exposure history}, which is in turn influenced by risk factors, geography, personal health choices, genetics, age, sex, \emph{etc.} Further variation stems from {\bf sequencing depth} (here, the total number of unique TCRs), collection methods introducing potential batch effects, and sequencing protocols. Based on past studies, T cell signatures responding to viral infection are estimated to account for $<1\%$ of the T cells in a repertoire~\citep{grifoni2020}. This means that only a small fraction of the $O(10^9)$ total unique TCR sequence set in a population of repertoires is relevant to a disease of interest.

Another challenge is that viral genomes often have high homology resulting in a significant number of {\bf shared antigens} \citep{Kieff1972}. The same is true for vaccines derived from specific sub-units of viral proteins. In both cases, we would expect predictive models to disentangle the shared TCRs from label-specific signals.
Finally, due to their fundamental role in antigen presentation~\citep{Reynisson2020} and TCR-antigen binding~\citep{Gruta2018} {\bf HLAs} directly influence an individual’s TCR repertoire. In some auto-immune diseases such as celiac disease and multiple sclerosis, HLAs can act as confounders because they directly influence both disease risk and the TCR repertoire~\citep{romanos2009analysis, Baranzini2011}. 
Our work focuses on demonstrating disease disentanglement and we leave HLA associations to future work.

\subsection{Enhanced Sequence Models}\label{sec:eslg}
 Since only a small fraction of TCRs are statistically significantly associated with a single disease, it is standard in immunomics modeling to perform Fisher's Exact Test (FET) per TCR. Specifically, $J$ TCRs, called {\bf{enhanced sequences}} (ES), are selected based on some $p$-value threshold, often reducing the feature set to numbers in the 100s or 1000s down from the {\em billions} in the full training set of repertoires.
 Given an ES list, a simple but robust logistic regression model can be trained using two features: the sum of ES counts in each repertoire and sequencing depth. Prior work has established \emph{Enhanced Sequence Logistic Growth} (ESLG) as a competitive benchmark for TCR-based diagnostics ~\citep{greissl2021immunosequencing,snyder2020magnitude}.  ESLG calibrates classification scores as a function of sequencing depth on the control population, which makes it generally more robust across depth. This simple model is suitable when we are learning from a single binary label but is ill-equipped to handle more complex immunomics data characterized by confounding or covariate shifts (such as age or HLA status) between cases and controls that may influence TCR selection. Moreover, ESLG assumes that the ES counts is only a function of the sequencing depth from the control population, which is restrictive and rarely satisfied in practice.

\subsection{Deep Learning Applied to TCR Sequences}
Our work focuses on statistical models of the discrete, count representation of TCR repertoires and we do not directly use the amino-acid sequence information.  However, there are several recent developments applying deep learning to TCR sequences (see~\citep{annurev-chembioeng} for an overview).  DeepTCR \citep{sidhom2021deeptcr} uses deep VAEs of TCR sequences to learn features useful for antigen binding prediction and repertoire classification using a multiple instance learning framework. DeepRC \citep{widrich2020modern} incorporates Hopfield networks into BERT transformers to classify repertoires in the context of existing metadata. Although promising, deep sequence models struggle to outperform simple regularized logistic regression models  \citep{kanduri2022profiling, emerson2017immunosequencing}.

\section{The Adaptive Immune Repertoire Invariant Variational Autoencoder}\label{sec:methods}



\begin{figure}[t]
	\centering



\begin{tikzpicture}
    \node[semifill={gray!50,ang=45}] (y) {$\Y^\l$};

    \node [node, below = of y] (z) {$\zv^\l_y$};
    \node [nodeobserved, left = of z] (x) {$\xv_j$};
    \node [node, left = of y] (zres) {$\zv_{\epsilonv}$};
    \plate [] {plate1} {(y)(z)} {$\d$};
    \plate [] {plate1} {(x)} {$J$};

    \draw [arrow] (y) -- (z);
    \draw [arrow] (z) -- (x);
    \draw [arrow] (zres) -- (x);
    \draw[->, dashed, bend right] (x) to (z);
    \draw[->, dashed, bend right] (x) to (zres);
    \draw[->, dashed, bend right] (z) to (y);
\end{tikzpicture}
	\caption{\textbf{Probabilistic graphical model for AIRIVA.} The generative and inference models are defined by solid and dashed arrows, respectively.  We assume $J$ TCRs $\xv_j$ in each repertoire along $K$ associated labels (observed or missing) $\Y^\l$, \emph{e.g.}, disease, sequencing depth, \emph{etc.}   We generate and infer the latent factors $\{\zv_{y}^{k}, \zv_{\epsilon} \}$ independently from other factors which leads to an interpretable disentangled representation, enabling counterfactual generation.  All distributions over latents $\zv$ are parameterized Normal distributions; TCR counts $\xv$ are generated by Poisson distributions; labels $\yv$ are distributed by Bernoulli, categorical, or log-normal depending on the setting.}
	\label{fig:pgm}
\end{figure}
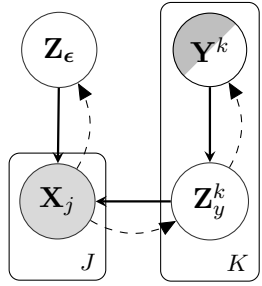

%
A variational autoencoder (VAE) is a probabilistic model that learns a mapping from some input data $\xv$, here corresponding to TCR repertoire counts, to a latent representation $\zv$~\citep{kingma_auto-encoding_2013,rezende2014stochastic}. This latent representation summarizes the information of the input data such that we can reconstruct the original data $\xv$ from the latent representation $\zv$ with high-fidelity. AIRIVA is based upon the Domain Invariant Variational Autoencoder (DIVA)~\citep{ilse2020diva} and the Capturing Characteristic VAE (CC-VAE)~\citep{joy2020capturing}, but applied and extended to immunomics. 
Both CC-VAE and DIVA leverage label information $\yv$ to disentangle the latent representation, where some of the dimensions of $\zv$ are constrained to also predict repertoire labels $\yv$ accurately. We can thus decompose $\zv$ as the concatenation of two kinds of latent variables. We call \textit{predictive latents} $\zv_{\yv}$ those latent dimensions that are label-specific, and \textit{residual latents} $\zv_{\epsilonv}$ those that are label-agnostic. Further, as illustrated in DIVA and CC-VAE, these models are capable of learning from samples with missing labels by marginalizing the unobserved labels.  In the immunomics setting this is very advantageous because it allows the model to learn from large cohorts of unlabelled repertoires. Figure~\ref{fig:pgm} shows the probabilistic graphical model of AIRIVA, which matches that of~\cite{ilse2020diva} and~\cite{joy2020capturing}. 
%


%


\subsection{Model Description and Objective Function}
    For simplicity and brevity, we restrict the AIRIVA objective function formulation to the {\bf fully-supervised} setting (see Appendices~\ref{app:ssdiva} and~\ref{app:ssccvae} for the {\bf semi-supervised} setting). 
    Our goal is to maximize the joint log-likelihood defined by the {\em intractable} integral over latent variables $\zv$:
\begin{align}
	\log p_{\thetav}(\xv , \yv) = \log  \int_{\zv}^{}  p_{\thetav}(\xv,  \yv, \zv) d\zv,
\end{align}
where $p_{\thetav}(\xv,  \yv, \zv) = p_{\thetav}(\xv | \zv) p_{\thetav}(\zv | \yv)$ and $\thetav$ represent the neural network parameters of the generative model.  By introducing a variational distribution  $q_{\phiv}(\zv | \xv)$, where $\phiv$ are the neural network parameters of the inference model, we can tractably maximize the variational evidence lower bound (ELBO):
\begin{align}
	\log p_{\thetav, \phiv}(\xv , \yv) & \ge \ \ell_{\text{ELBO}}(\thetav, \phiv;  \xv, \yv)\,
\end{align}
By factorizing both the generative and inference model over $K$ labels, we have:
\begin{align}
\label{eq:prior}
p_{\thetav}(\zv | \yv) &= p(\zv_{\epsilonv}) \prod_{\l} 	p_{\thetav}^\l (\zv_{\yv}^\l | \yvs^\l)\, \\
\label{eq:posterior}
q_{\phiv}(\zv | \xv) &= q_{\phiv}^{\epsilonv}(\zv_{\epsilonv} | \xv) \prod_{\l} 	q_{\phiv}^\l (\zv_{\yv}^\l | \xv).
\end{align}
Following the formulation from \citet{ilse2020diva}, the ELBO simplifies as follows:
\begin{align}
    \ell_{\text{ELBO}}(\thetav, \phiv ;  \xv, \yv) & = 
    \EE_{q_{\phiv}(\zv| \xv)}\left[ \log  p_{\thetav}(\xv | \zv) \right] \nonumber \\
	& - \beta_{\epsilonv} \cdot KL\left( q_{\phiv}^{\epsilonv}(\zv_{\epsilonv}| \xv)  \, || \,  p(\zv_{\epsilonv})\right)  \nonumber \\
    & - \sum_{\l} \beta_\l \cdot KL \left ( q_{\phiv}^\l (\zv_{\yv}^\l | \xv) \, || \, p_{\thetav}^\l (\zv_{\yv}^\l | \yvs^\l) \right), \label{eq:elbo}
\end{align}
where $\{ \beta_{\epsilonv}, \beta_{\l} \}$ are additional hyperparameters that can be scheduled during training to prevent posterior collapse by KL divergence annealing \citep{higgins2016beta, bowman2015generating} or otherwise set to trade-off various parts of the objective function. 

This objective function is missing a key driver required to enforce disentanglement, a label classification term~\citep{locatello2019challenging, kim2018disentangling}.  Following~\cite{ilse2020diva}  
we introduce an auxiliary objective:
\begin{align}
	\ell_{\text{aux}}(\omegav, \phiv;  \xv, \yv) \!=\! \sum_{\l} \!\alpha_\l  \EE_{q_{\phiv}^\l (\zv_{\yv}^\l | \xv)} \left[ \log q_{\omegav}^\l (\yvs^\l | \zv_{\yv}^\l)\right]\!,
\end{align}
with independent 
label-specific prediction neural networks with parameters $\omegav$ to encourage learning of label-specific independent factors\footnote{Using a similar graphical model, \cite{joy2020capturing} derives the objective assuming $q_{\phiv}(\zv | \xv, \yv)$, rather than $q_{\phiv}(\zv | \xv)$ in \cite{ilse2020diva},  resulting in the classifier naturally appearing in the objective.  However, the final objective in both cases is similar, modulo a weighting in the expectation.} and $\alpha_\l$ hyperparameters for weighting the auxiliary predictor. 
For the {\bf fully supervised} setting, the complete AIRIVA regularized variational objective for  jointly learning all model parameters $\{\thetav, \psiv, \omegav\}$  via stochastic gradient ascent is:
\begin{align}
	\ell( \thetav, \phiv, \omegav; \xv, \yv)\!=\!  \ell_{\text{ELBO}}(\thetav, \phiv ; \xv, \yv)\! +\!	\ell_{\text{aux}}(\omegav, \phiv; \xv, \yv). \label{eq:optfunction}
\end{align}

\subsection{In-silico Generation of Repertoires}\label{sec:counterfactuals}
Following a standard VAE setup, we generate synthetic (in-silico) repertoires $\xv \sim p_{\thetav} (\xv| \zv)$ from AIRIVA, where $\zv$ is drawn according to the prior distribution from Eq. \eqref{eq:prior} also known as {\bf conditional generation}. With conditional generation, we can synthesize samples that match the \emph{factual} (empirical) TCR repertoire distribution. This can be particularly useful to balance training datasets, by increasing the number of repertoires for minority subgroups via targeted data augmentation. However, in the context of immunomics, we may want to answer \emph{what if} questions such as: ``\textit{what would the immune response look like if the patient had not been exposed to certain antigens}'', or ``\textit{which set of TCRs would we expect to observe if the healthy control had been vaccinated?}''. So motivated, we leverage the factorized posterior from Eq.~\eqref{eq:posterior} and conditional prior in Eq.~\eqref{eq:prior} for {\bf counterfactual generation}. Below we detail the procedure to generate counterfactuals and infer label-specific TCRs.

\paragraph{Counterfactual Generation}
Following the potential outcomes framework~\citep{rubin2005causal}, we assume a binary \emph{treatment} $Y^{k}$ where each TCR $X_{j}$ has two potential outcomes denoted by $X_{j}(Y^{k}=0)$ and $X_{j}(Y^{k}=1)$. In practice, for each sample, we only observe the \emph{factual} outcome $X_{j}(Y^{k})$, while the \emph{counterfacual} $X_{j}(1-Y^{k})$ is unobserved. Hence, we generate counterfactuals with AIRIVA as follows:
\begin{enumerate}
\itemsep0em 
    \item sample from posterior $\zv \sim q_{\phiv}(\zv | \xv)$ in Eq.~\eqref{eq:posterior} (equiv. to computing exogenous noise).
    \item assign posterior \emph{label-specific} latent factor $\zv_{\yv}^k$ to the sample from the conditional prior in Eq.~\eqref{eq:prior} $\zv_Y^k \sim p_{\thetav}(\zv_{\yv}^k | 1-Y^{k})$ (equivalent to \emph{intervening} on $Y^{k}$ since $\zv_{\yv}^k$ is independent to all the other latent factors by construction).
    \item sample \emph{counterfactual} immune repertoire $\xv (1-Y^{k}) \sim p_{\thetav}(\xv | \zv)$.
\end{enumerate}


\paragraph{Inferring Label-Specific TCRs}
\label{sec:cate}

Finally, we leverage generated counterfactuals from AIRIVA to infer label-specific TCRs. To quantify the expected effect of a given label $Y^k$ on each individual TCR $X_j$ for a given subpopulation, we can compute the \textit{conditional average treatment effect} (CATE), defined as:
\begin{equation}
\text{CATE}_j(\y^{\setminus k}) \defeq \mathbb{E}_{X_j}\left[ X_{j}(Y^{k}=1) - X_{j}(Y^{k}=0) \given \yv^{\setminus k}=\y^{\setminus k} \right],
\end{equation}
where a subpopulation is defined by a fixed value assignment for the other labels $\yv^{\setminus k}=\y^{\setminus k}$. In our experiments, we use the $\text{CATE}_j$ to rank TCRs s.t. the inferred label-specific TCRs have the highest $\text{CATE}_j$.

\section{Experiments}\label{sec:results}

In this section, we demonstrate the capacity of AIRIVA to disentangle disease-specific TCRs in the context of two infectious diseases: COVID-19 and the Herpes
Simplex Virus. We also show that AIRIVA can learn from unlabelled samples, generate realistic in-silico TCR repertoires, and learn a latent space that aligns with biological intuition.  
We first briefly describe general procedures for training, model selection and model evaluation used in our experiments. This is then followed by detailed case-study experimental procedures and results.  See Appendix for further experimental details, results and analyses.

\subsection{Experimental Details}\label{sec:exp-setup}
Before training, we rank TCRs using one-sided FET $p$-values on occurence in cases vs. controls and select the top TCRs. This ensures selection of public TCRs \citep{emerson2017immunosequencing}. {\bf Training} consists of stochastic gradient ascent using the objective function defined by Eq.~\eqref{eq:optfunction}.
Because the objective function is composed of several (hyperparameter-weighted) sub-optimization problems, the typical {\bf model selection} procedure of choosing the model with the best maximum objective on the validation set sometimes results in poor models, \emph{e.g.}, that have good reconstruction but have ignored the latent factor constraints.  We have empirically found that a uniform mixture of classification metrics\footnote{We use {\em concentrated} area under the ROC curve (AUCROC)~\citep{Swamidass2010} which modifies typical AUROC by emphasizing sensitivity.} on the factual (validation) \textit{and} on counterfactual (validation) data reliably selects good models. This aligns with previous works that link counterfactual performance to disentanglement and model robustness~\citep{suter2019robustly, besserve2018counterfactuals}.
We are interested in {\bf evaluating} the following key aspects of AIRIVA: (1) predictive performance of AIRIVA when compared to ESLG (see Section~\ref{sec:eslg}) and a standard multi-label feed forward network (FFN) classifier; (2) qualitively inspect latent factor disentanglement,  both from its conditional prior and posterior distributions; (3) similarly evaluate counterfactuals in the latent space for correctness; (4) when applicable, validate AIRIVA estimated TCR-associations based on CATEs (see Section~\ref{sec:cate}) with an independently verified set of annotated TCRs from an external assay dataset (see Section~\ref{sec:mira}). 
\subsection{COVID-19 Case Study}

The viral envelope of SARS-CoV-2 is made up of four proteins: the membrane protein, the envelope protein, the nucleocapsid protein and the spike protein~\citep{jackson2022}. The spike protein is believed to be responsible for binding to host cells and initiating viral infection, and has therefore been the target of vaccine development~\citep{mccarthy_covid2021}. During natural infection, we expect to see the expansion of TCRs associated with any one of these proteins, while in the case of vaccination\footnote{We consider viral vector vaccines AZD1222 (Oxford-AstraZeneca) and Ad26.COV2.S (Johnson \& Johnson), as well as mRNA vaccines BNT162b2 (Pfizer) and mRNA-1273 (Moderna).} we should only observe the expansion of TCRs binding the spike protein. We would like to build a model that can disentangle natural infection from vaccination. Because TCRs clonally expanded during natural infection are a superset of the spike-only TCRs expanded following vaccination, disentangling both sets of TCRs is a challenging task.

\paragraph{Data Cohorts}
To train our models we used $1,954$ samples from donors with natural SARS-CoV-2 infection, $477$ healthy donors post SARS-CoV-2 vaccination, $5,198$ healthy controls sampled prior to March 2020, and $0$ donors who were naturally infected and later vaccinated. The classifiers are tested on a holdout set of $525$ samples from naturally infected donors, $400$ vaccinated donors, $100$ donors who were naturally infected and later vaccinated, and $4,606$ healthy controls.

We posit $\yv = \{\yvs_\text{ns}, \yvs_\text{s}, \yvs_\text{depth}\}$, where $\yvs_\text{ns}$ and $\yvs_\text{s}$ refer to exposure to the non-spike and spike proteins respectively, and $\yvs_\text{depth}$ refers to sequencing depth---a standardized measure of log total template count---as there is a strong depth imbalance between the vaccinated subpopulation and others. For $\yvs_\text{ns}$, naturally infected samples are positives, and vaccinated and healthy controls are negatives. For $\yvs_\text{s}$, both naturally infected and vaccinated samples are positives and healthy controls are negatives.


\paragraph{TCR Selection}
We ran a sweep of 100 ESLG models for each prediction task (spike and non-spike) to determine the best threshold number $\eta$ of input TCRs, where $\eta \in [200, 500, 1000, 5000, 10000]$. Further, we select $\eta$ that maximizes the predictive performance of ESLG in the validation set. The final input TCRs used to train AIRIVA are based on: $i)$ 1,254 TCR sequences from the union of 1,000 most significant $\eta$ TCRs selected by FET per binary label (non-spike and spike); and $ii)$ 342 TCRs associated with cytomegalovirus (CMV) to further test AIRIVA's robustness to noise induced by an unrelated signal from another infectious disease, yielding a total of $J=1,596$ input TCRs.

\paragraph{External Spike and Non-Spike Annotations}\label{sec:mira}
In addition to repertoire-level labels, in the case of COVID-19, we have access to in-vitro labels for spike and nonspike TCR associations from external MIRA (\textbf{M}ultiplexed \textbf{I}dentification of T cell \textbf{R}eceptor \textbf{A}ntigen specificity) assay data~\citep{klinger2015multiplex}.
This data comprises over 400k experimentally-derived TCR-antigen binding pairs (\emph{hits}) for the spike protein, and 1.2M hits for non-spike proteins~\citep{nolan2020large}. A total of 291 spike and 462 non-spike MIRA hits appear in our input set of 1,596 TCRs. By intersecting public COVID-19-associated TCRs with this database, it is possible to associate a subset of the TCRs to an antigen (spike or non-spike) with high confidence \citep{Li2022}.
We use these spike and non-spike protein-specific TCRs from MIRA to assess the quality of the inferred TCR-label associations from AIRIVA.
%
\subsubsection*{Covid-19 Results}
\paragraph{Disease Classification}
FET sequence selection for the non-spike label uses the natural infection subgroup as cases, however, these contain both the spike and non-spike signal. Hence, we expect non-spike AIRIVA predictions to outperform predictions from the non-spike ESLG model given natural infection samples as cases and vaccinated samples as controls. This is due to ESLG's non-robustness to noisy input TCRs (\emph{i.e.}, shared TCRs across disease labels).
To demonstrate this, we report holdout performance on the subgroups listed in Table~\ref{tab:subgroups}.

\begin{table}[htb!]
    \centering
    \caption{\textbf{Subgroups used for evaluation of COVID-19 models.}}
    \label{tab:subgroups}
        \begin{tabular}{rcc}
    \toprule
    &  \textbf{Cases} &  \textbf{Controls}\\
    \midrule
    \textbf{Overall}     & all natural infection    & vaccinated, healthy controls        \\
    \textbf{Unvaccinated}   & natural infection    & healthy controls       \\
    \textbf{Vaccinated}   & natural infection + vaccinated    & vaccinated       \\
    \bottomrule
    \end{tabular}
\end{table}

\begin{figure}[th]
    \centering
    \begin{minipage}{0.32\textwidth}
        \centering
        \includegraphics[width=1\textwidth]{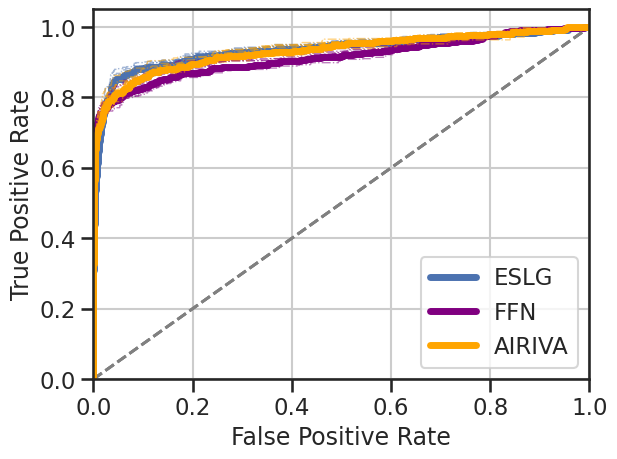}
        \caption*{(a) Overall}\label{fig:roc_covid}
    \end{minipage}
    \begin{minipage}{0.32\textwidth}
        \centering
        \includegraphics[width=1\textwidth]{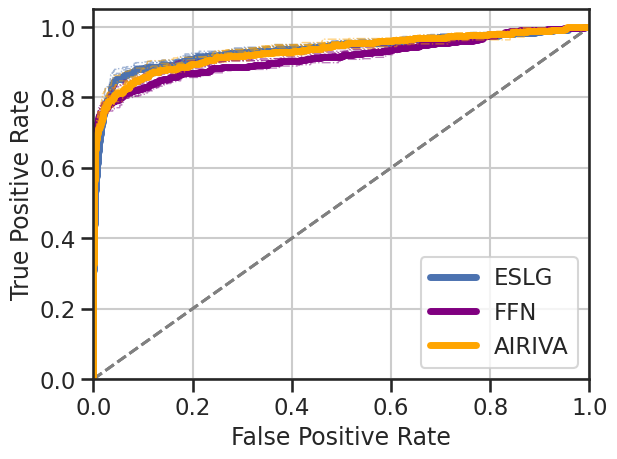}
        \caption*{(b) Unvaccinated}\label{fig:roc_covid_0}
    \end{minipage}
    \begin{minipage}{0.32\textwidth}
        \centering
        \includegraphics[width=1\textwidth]{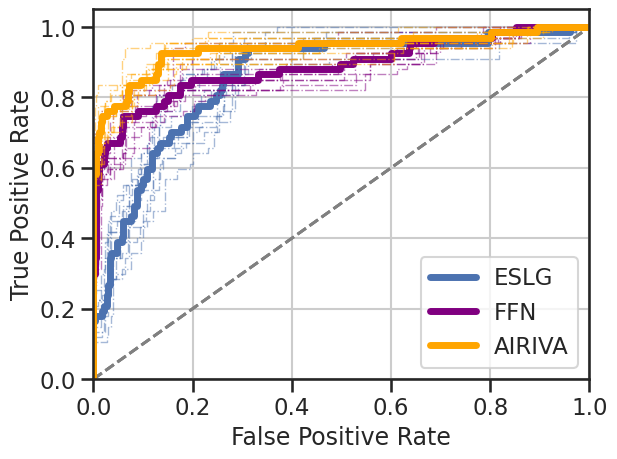}
        \caption*{(c) Vaccinated}\label{fig:roc_covid_1}
    \end{minipage}
     \caption{\textbf{ROC curves for non-spike disease models.} Bold lines correspond to the ROC curve on holdout; colored dashed lines correspond to 10 additional bootstrap samples. ESLG performs poorly for the vaccinated subgroup, while AIRIVA performance remains competitive, with similar predictive performance compared to the overall population.}
     \label{fig:roc_covid_all}
\end{figure}

Table~\ref{tab:performance_covid} compares the discriminative value of non-spike AIRIVA against FFN and ESLG for these different case/control groups.\footnote{In the Appendix, we report comparative performance for the spike label; AIRIVA achieves similar performance against ESLG and FFN, which is expected and further validates the modeling framework.} We can see that for the entire population as well as when comparing natural infection to healthy controls all three models perform comparably. However, as expected, AIRIVA significantly outperforms ESLG and FFN when comparing natural infection samples with subsequent vaccination against vaccinated samples. Figure~\ref{fig:roc_covid_all} shows the corresponding Receiver-Operating  Curves (ROC) for the three groups considered, highlighting the degradation in model performance of ESLG for the vaccinated subgroup.
These results indicate that AIRIVA is robust to shared antigens in disease labels and can disentangle spike-associated disease signal from non-spike signal.

\begin{table}[t]
    \centering
    \scriptsize
    \caption{\textbf{Comparison of non-spike disease models}. Sensitivity at 98\% specificity and AUROC for predicting natural infection, overall and stratified by vaccination status (in both cases and controls), using 100 bootstraps samples. AIRIVA performs comparatively against Enhanced Sequence Logistic Growth (ESLG) overall, but is more robust at separating naturally infected samples with subsequent vaccination from vaccinated samples.}
    \label{tab:performance_covid}
    \begin{tabular}{ccccccc}
        \toprule
        & \multicolumn{2}{c}{\textbf{Overall} } & \multicolumn{2}{c}{\textbf{Unvaccinated} } &
      \multicolumn{2}{c}{\textbf{Vaccinated} } \\
        \cmidrule(lr){2-3}
        \cmidrule(lr){4-5}
        \cmidrule(lr){6-7}
        Model &  Sensitivity & AUROC  & Sensitivity & AUROC & Sensitivity & AUROC\\
        \midrule
        ESLG & 0.74 $\pm$ 0.07 & 0.94 $\pm$ 0.02 & 0.85 $\pm$ 0.04 & 0.94 $\pm$ 0.02 & 0.27 $\pm$ 0.15 & 0.86 $\pm$ 0.06 \\
        FFN & 0.76 $\pm$ 0.05 & 0.92 $\pm$ 0.02 & 0.77 $\pm$ 0.05 & 0.92 $\pm$ 0.02 & 0.63 $\pm$ 0.16 & 0.89 $\pm$ 0.06 \\
        AIRIVA & 0.76 $\pm$ 0.06 & 0.93 $\pm$ 0.02 & 0.76 $\pm$ 0.05 & 0.93 $\pm$ 0.02 & 0.73 $\pm$ 0.12 & 0.94 $\pm$ 0.04 \\
        \bottomrule
    \end{tabular}
\end{table}

\begin{figure}[t]
    \begin{minipage}[b]{0.52\linewidth}
	\centering
         \includegraphics[width=1\textwidth]{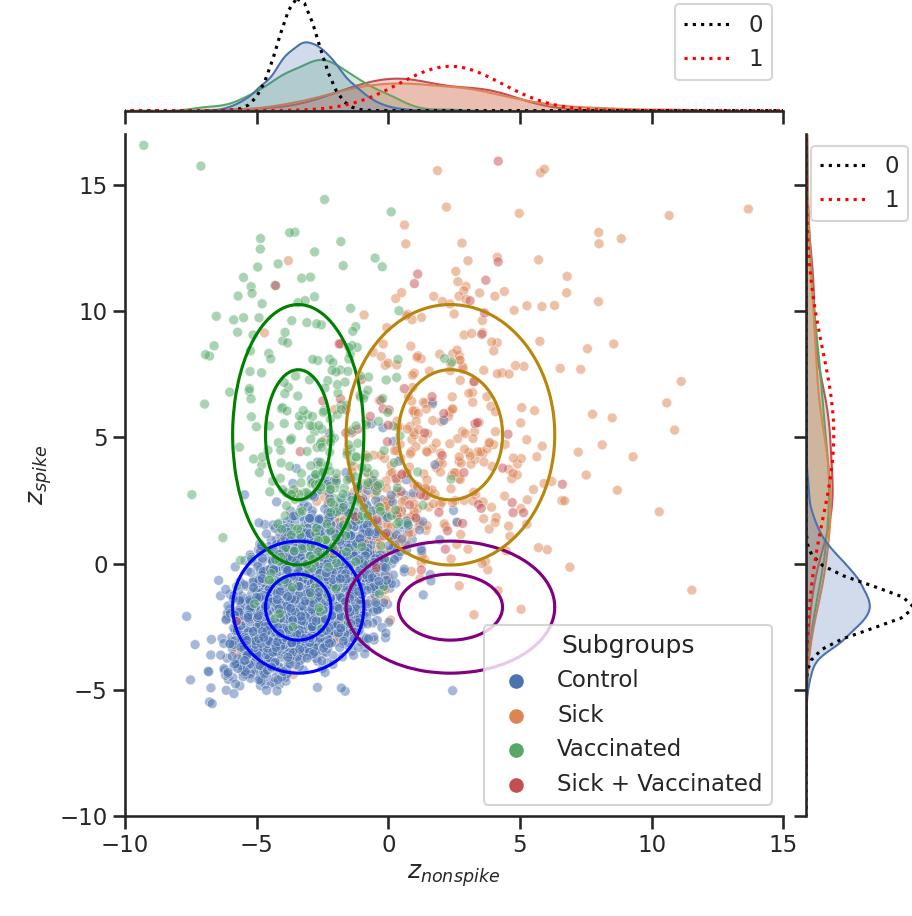}
        \caption*{(a) Stratified by subgroup}
    \end{minipage}
    \begin{minipage}[b]{0.4\linewidth}
        \centering
            \includegraphics[width=\textwidth]{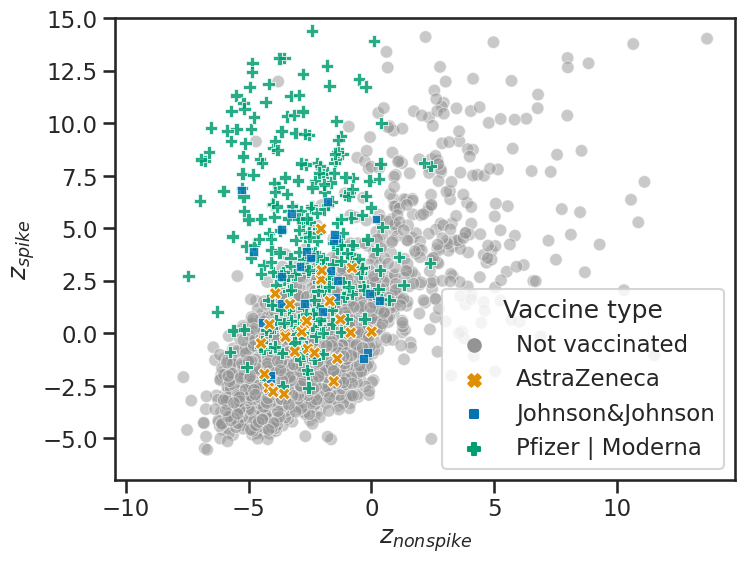}
            \includegraphics[width=\textwidth]{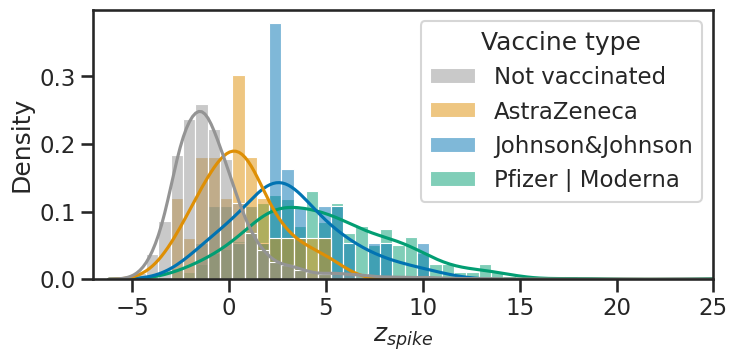}
            \caption*{(b) Stratified by vaccine type}
            \label{fig:vaccine_latent_space}
    \end{minipage}
	\caption{\textbf{Latent space $q_{\phiv}(\zv_{\yv}|\xv)$ for COVID-19 in holdout}. AIRIVA learns an interpretable latent space that matches biological intuition. (a) 
 Each point corresponds to one posterior sample. Conditional priors $p_{\thetav}(\zv_{\yv}|\yv)$ for each combination of labels are represented as ellipses, each circle denoting one standard deviation. On the sides, we show the marginal posteriors as continuous lines and conditional priors as dashed lines. (b) Top: posterior samples for vaccinated subgroups (colored),
    samples for other subgroups are shown in grey; Bottom: histogram of posterior samples on the $Z_\text{spike}$ factor dimension.}\label{fig:latent_space_COVID}
\end{figure}

\paragraph{Interpreting Learnt Latent Space} 
Figure~\ref{fig:latent_space_COVID}(a) plots posterior samples from $q_{\phiv}(\zv_{\yv}|\xv)$ for the holdout data along with the exact learned prior predictive distributions $p_{\thetav}(\zv_{\yv}|\yv)$ in the latent subspace ($ \zv_{\yv} =\{Z_\text{non-spike}$, $Z_\text{spike}$\}), stratified by subgroups. Note that one of the label groups, corresponding to ``\textit{non-spike only}'' exposure, is never observed in our data since both spike and non-spike co-occur in natural infection. As expected, the vaccinated subgroup exhibits only spike signal (top-left), whereas COVID-positive cases,with and without vaccination, show exposure to both spike and non-spike antigens (top-right).

Such a structured latent space provides a way to interpret both the composition and the strength of the observed immune response and could elicit new biological insights.
For example, Figure~\ref{fig:latent_space_COVID}(b) shows posterior samples of the vaccinated-only repertoires in holdout stratified by vaccine type; samples from other subgroups are plotted in grey. 
Our model suggests that repertoires that were administered the AstraZeneca viral vector AZD1222 vaccine exhibit lower spike T-cell response compared to mRNA vaccines from Johnson \& Johnson, Pfizer, or Moderna which is consistent with findings in the literature \citep{Schmidt2021,Prendecki2021,Marking2022}.

\paragraph{Conditional Generation}
To demonstrate the disentanglement of spike and non-spike TCR response in AIRIVA, we evaluate whether conditionally generated repertoires have the characteristics we expect.
Specifically, repertoires of vaccinated and COVID-positive individuals should contain higher counts of spike-specific TCRs, while non-spike-specific TCRs should only activate for COVID-positive individuals. To test this, we generate repertoires conditionally by intervening on the spike and non-spike labels, at a fixed mean repertoire sequencing depth.
%
\begin{figure}[th]
\centering
\includegraphics[width=\linewidth]{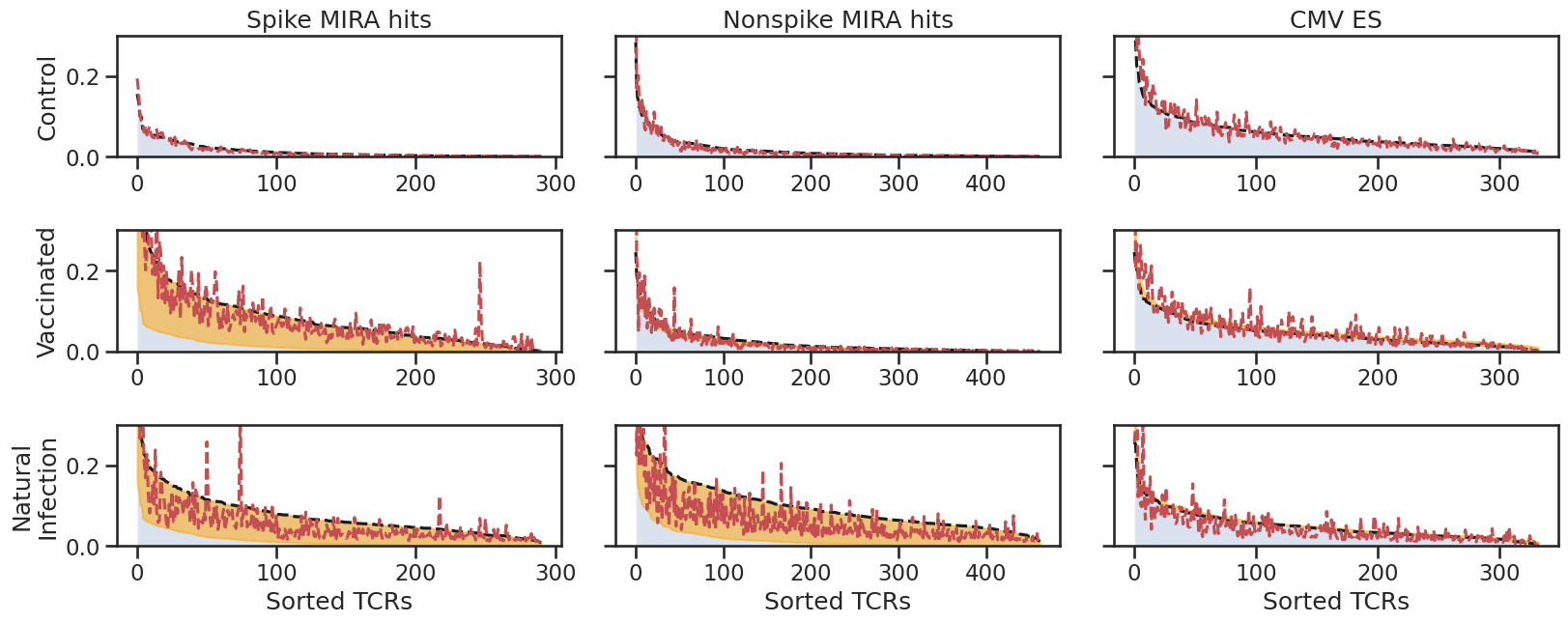}
\caption{\textbf{Conditional generation with AIRIVA}. Per-TCR mean counts estimated by AIRIVA (average Poisson rate as dashed black line) or empirically (dashed red line), averaged across repertoires. We generate 5,000 in-silico repertoires for each subgroup; the pool of TCRs is the same as the initial set of 1596 input TCRs for all subgroups. TCRs are stratified according to TCR annotations and sorted in descending order per estimated CATE. The yellow area highlights the difference w.r.t controls. In-silico repertoires generated by AIRIVA exhibit per-TCR mean counts similar to the observational data.
}\label{fig:cond_generation}
\end{figure}


Figure~\ref{fig:cond_generation} shows the average count per TCR estimated by AIRIVA (\emph{i.e.}, per-TCR Poisson rate for conditionally generated repertoires, averaged across repertoires) and empirical average count per TCR. We split TCRs based on external annotations of spike and non-spike MIRA hits (described in Subsection~\ref{sec:mira}), and CMV enhanced sequences.
As expected, generating ``vaccinated" repertoires with spike only yields increased average per-TCR count for spike MIRA hits. Moreover, generating repertoires with natural infection (conditioning on positive spike and non-spike labels) shows increased average per-TCR counts in both spike and non-spike MIRA TCRs. Average count of CMV-associated sequences remains negligible across all label groups. Thus providing empirical evidence that AIRIVA is robust to irrelevant background signal which is captured by the \emph{label-agnostic} latent factor $\zv_{\epsilonv}$. The conditionally generated TCR counts match factual counts well in the controls, however, AIRIVA overestimates counts in the natural infection samples. One hypothesis for this behavior is that sequence occurrence in controls is mostly due to non-expanded naive TCRs which occur randomly in repertoires in accordance with a well-defined generative process. This sampling should match our assumed Poisson distribution of TCR counts well. However, TCR counts in subpopulations that contain expanded TCRs may not follow a Poisson distribution because of extra variance introduced in the TCR counts by clonal expansion.   

%

\paragraph{Consistent Counterfactual Generation for COVID-19}

\begin{figure}[]
        \centering
        \includegraphics[width=1.0\linewidth]{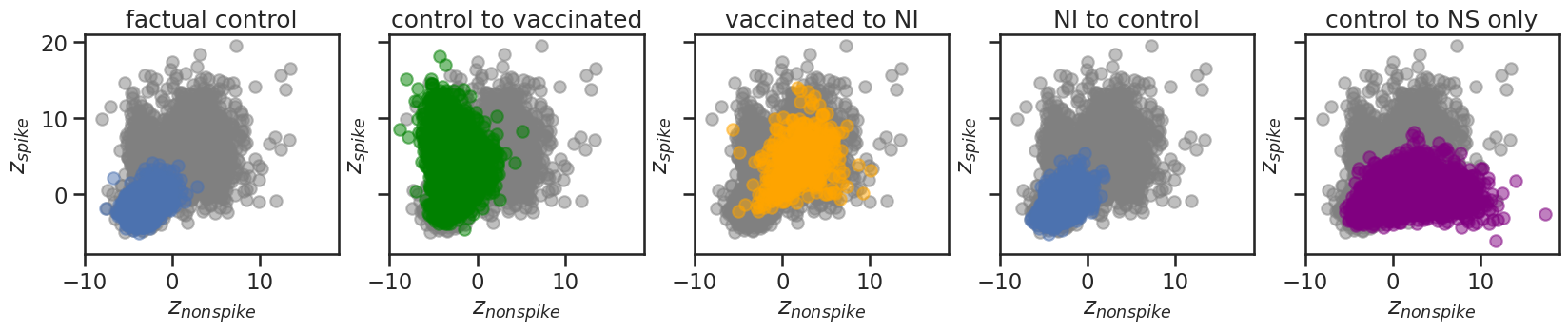}
        \includegraphics[width=1.0\linewidth]{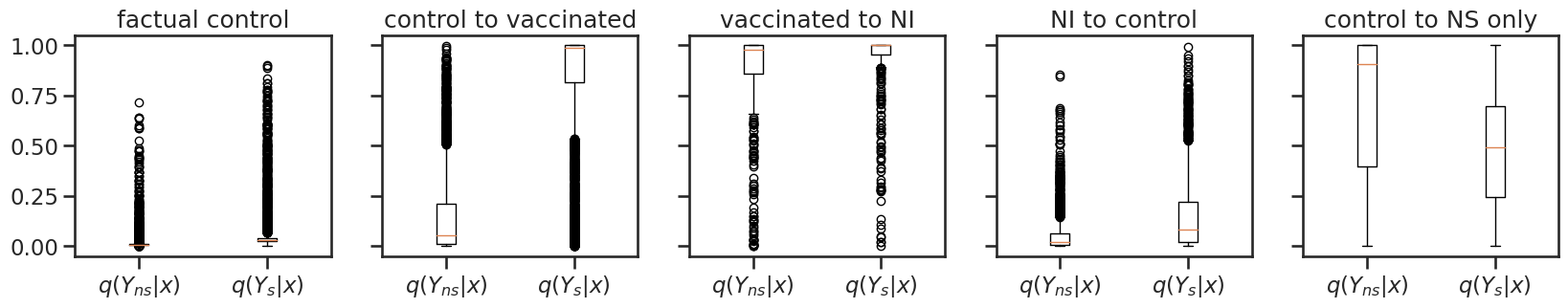}
	\caption{\textbf{Counterfactual generation for COVID-19}. (First row) Posterior samples $\zv_{\yv} \sim q_{\phiv}(\zv_{\yv}|\xv)$ for factual data (first column) and counterfactual data (other columns). (Second row) Posterior predictive samples $\yv \sim q_{\omega}(\yv \given \zv_{\yv})$ for each binary label. See the Appendix for all possible counterfactual combinations stratified by subgroups. 
	}
	\label{fig:counterfactuals_main}
\end{figure}

Figure~\ref{fig:counterfactuals_main} shows posterior samples (first row) and posterior predictive samples (second row) for one of the observed subgroups (first column), and counterfactuals (other columns) transformed from one subgroup to another.
%
This figure demonstrates that we can generate consistent counterfactuals that exhibit expected properties by manipulating specific labels. Importantly, the latent representation of such counterfactuals maps to a similar region as their corresponding factual subgroups in Figure~\ref{fig:latent_space_COVID}. Note that the last column corresponds to an \emph{unobserved} repertoire subgroup: these are out-of-distribution counterfactuals. Interestingly, AIRIVA captures the correct direction of variation in the latent space for that novel subgroup without introducing additional inductive biases often necessary for zero-shot learning.

\paragraph{Inferring Label-Specific TCRs}

\begin{figure}
    \centering
    \includegraphics[width=\linewidth]{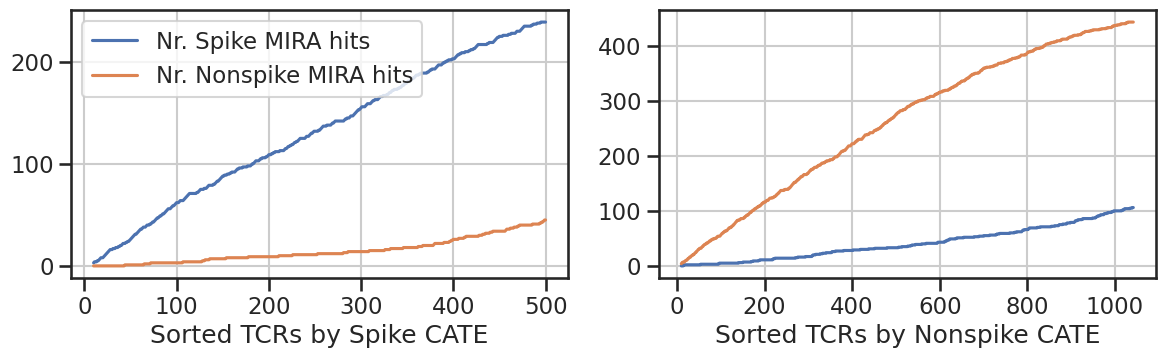}
    \caption{\textbf{Identification of Spike and Non-spike specific TCRs with CATE.} We show the proportion of spike and non-spike MIRA TCRs recovered when traversing spike and non-spike CATEs in descending order (only TCRs with MIRA annotations are considered). The x-axis for both plots corresponds to the index where all 291 spike and 462 non-spike MIRA hits are recovered (the maximum value of the y-axes).}
    \label{fig:my_label}
\end{figure}

We use AIRIVA to explicitly infer label associations for input TCRs, by evaluating CATEs for spike and non-spike exposure respectively per TCR, as described in Section~\ref{sec:cate}. Specifically, we define Spike CATE and Non-spike CATE for each TCR $X_j$ as follows:
\begin{eqnarray}
\label{eq:spike_cate}
\text{Spike CATE} & \defeq & \mathbb{E}_{X_j}\left[ X_{j}(Y_{\text{s}}=1) - X_{j}(Y_{\text{s}}=0) \given Y_{\text{ns}}=0 \right] \nonumber \\ 
\label{eq:non_spike_cate}
\text{Non-spike CATE} & \defeq & \mathbb{E}_{X_j}\left[ X_{j}(Y_{\text{ns}}=1) - X_{j}(Y_{\text{ns}}=0) \given Y_{\text{s}}=1 \right],
\end{eqnarray}
where each potential outcome is either observed (factual) or estimated via counterfactuals generated by AIRIVA \footnote{Note that we condition on Spike-positive for Non-spike CATE as both source and target subgroups in counterfactual generation need to be observed for counterfactuals to be well-defined, as discussed in Section~\ref{sec:cate}}. 
We rank COVID-19 TCRs in descending order according to their spike or non-spike CATE values. Then we leverage ground truth MIRA labels to compute a cumulative count of spike or non-spike MIRA TCRs (illustrated as blue or orange, respectively in Figure~\ref{fig:my_label}).
We expect TCRs with the largest spike CATE to be mostly spike MIRA TCRs and vice-versa which is indeed what we observe. Note that we are only able to assign a little over half of our total TCRs from FET to spike or non-spike via MIRA and are not showing unassigned TCRs. 
%
We evaluate AIRIVA's ability to recover ground truth spike on non-spike MIRA sequences using CATEs \eqref{eq:spike_cate} via average precision and false discovery rates (FDR), defined as:
%
\begin{eqnarray}
\text{average precision} = \frac{1}{J} \sum_{j} \frac{{\rm TP}_j}{ {\rm TP}_j + {\rm FP_j }} \nonumber \\
\text{average FDR} = 1 - \text{average precision}\,.
\end{eqnarray}
%
To compute the \emph{average spike precision rate}, we assign true positives (TP) to the computed cumulative spike MIRA hits and false positives (FP) to cumulative non-spike MIRA hits at TCR index $J$. Similarly, for the \emph{average non-spike precision rate} we assign TP and FP to the cumulative non-spike and spike MIRA hits, respectively. Table~\ref{tab:fdr+precision} reports average precision and FDR for TCRs ranked according to the spike and non-spike CATE for index values $J = 10, \ldots, 100$. We can interpret the table as a confusion matrix for spike and non-spike hits: we want high values on the diagonal (precision), and low values in off-diagonal elements (FDR). Indeed, we report a high precision of $0.95$ and $0.91$ for spike and non-spike predictions, respectively. For instance, this implies that $95\%$ of our spike sequence assignments are correct. Conversely, we report low FDRs for both spike and non-spike predictions. Finally, as expected we report low spike and non-spike FDRs for CMV TCRs. 


\begin{table}[htb!]
    \centering
    \caption{\textbf{Percentage of each type of TCR among top-TCRs ranked by AIRIVA.} We report fraction of Spike MIRA hits, Non-spike MIRA hits, and CMV Enhanced Sequences among all annotated TCRs present in the top $J$ as ranked by Spike CATE and Non-spike CATE, averaging over $J$ between 10 and 100.}
    \label{tab:fdr+precision}
        \begin{tabular}{rccc}
    \toprule
    &  \textbf{Spike hits} &  \textbf{Non-spike hits} & \textbf{CMV ES}  \\
    \midrule
    \textbf{Spike  CATE}     & 0.95    & 0.05    & 0.0    \\
    \textbf{Non-spike CATE}   & 0.09    & 0.91    & 0.06    \\
    \bottomrule
    \end{tabular}
\end{table}

\subsection{HSV Case Study}
HSV-1 and HSV-2 are two members of the Herpes Simplex Virus (HSV) family. The two viruses share approximately 80\% of their genome~\citep{greninger2018ultrasensitive}, and will therefore have partially overlapping antigen sets and corresponding TCRs, making it challenging to build a model which accurately disentangles their labels.
Infections for both are frequent in the population, with $\approx$ 67\% prevalence for HSV-1 and $\approx$ 13\% prevalence for HSV-2~\citep{james2020herpes}. Infection status for either virus is independent of the other and infections frequently co-occur. 

\paragraph{Data Cohorts}
Our dataset consists of samples labeled for both HSV-1 and HSV-2 using multiplexed immunoglobulin G (IgG) serology. The dataset contains 1,125 repertoires in the training set, and 289 in the holdout set. We are unable to confidently assign a label for HSV-1 and/or HSV-2 in roughly 20\% of our samples. We include these unlabeled samples for AIRIVA in a semi-supervised context (SS-AIRIVA). Here we build an AIRIVA model to disentangle the two labels. We posit the set of labels $\yv = \{\yvs_\text{hsv1}, \yvs_\text{hsv2}\}$.

\paragraph{TCR Selection}

We ran a sweep of 100 ESLG models for each prediction task (HSV-1 and HSV-2) with a $p$-value threshold for FET, where
$p \in [0.0001, 0.005, 0.001, 0.05]$ and selected $p$ that maximizes ESLG predictive performance in the validation set. Based on this criteria, we selected a $p$-value threshold of 0.001 for the FET-based TCR selection step, yielding a total of 158 input TCRs.
\subsubsection*{HSV Results}
\begin{figure}[th]
    \centering
    \begin{minipage}{1.0\textwidth}
        \centering
        \includegraphics[width=0.32\textwidth]{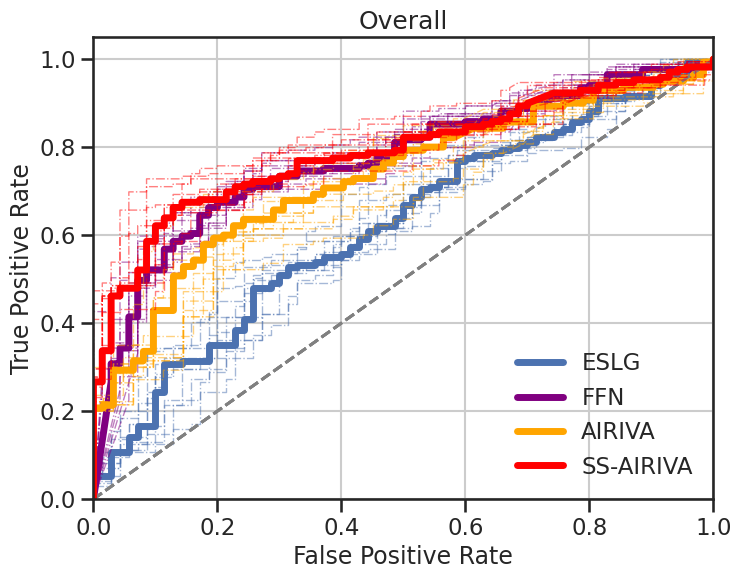}
        \includegraphics[width=0.32\textwidth]{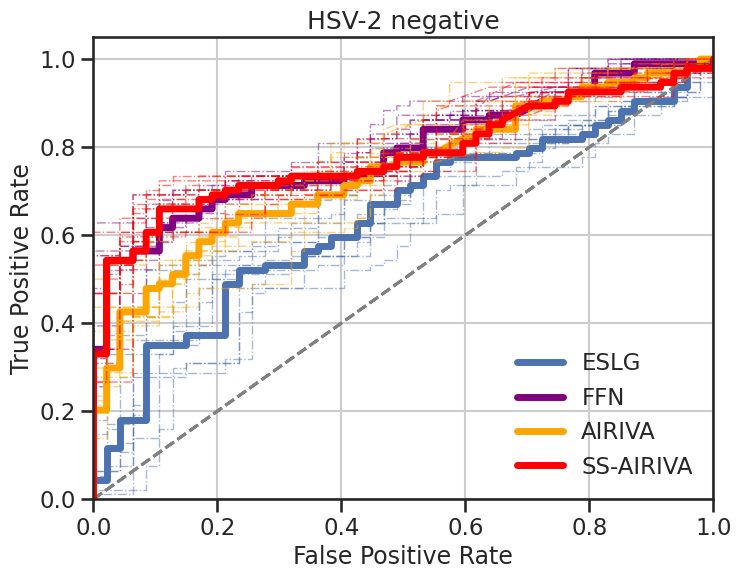}
        \includegraphics[width=0.32\textwidth]{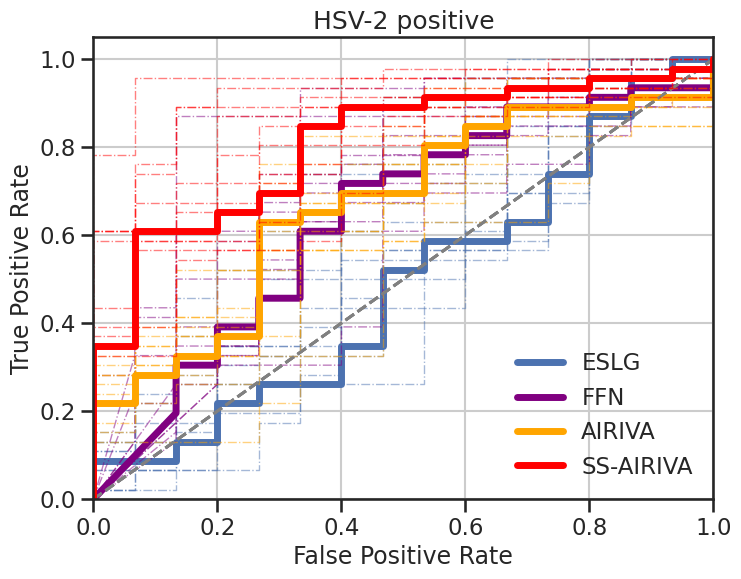}
        \caption*{(a) HSV-1 Prediction Task}\label{fig:hsv1}
    \end{minipage}
    
    \vspace{0.3cm}
    \begin{minipage}{1\textwidth}
        \centering
        \includegraphics[width=0.32\textwidth]{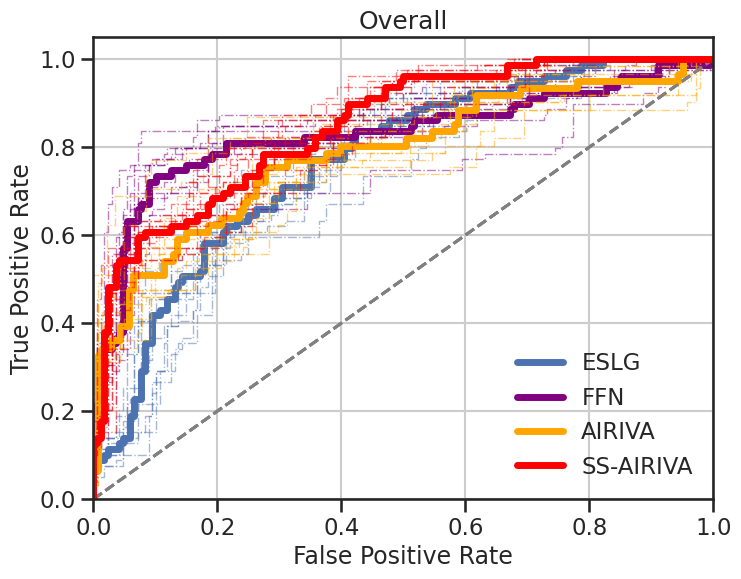}
        \includegraphics[width=0.32\textwidth]{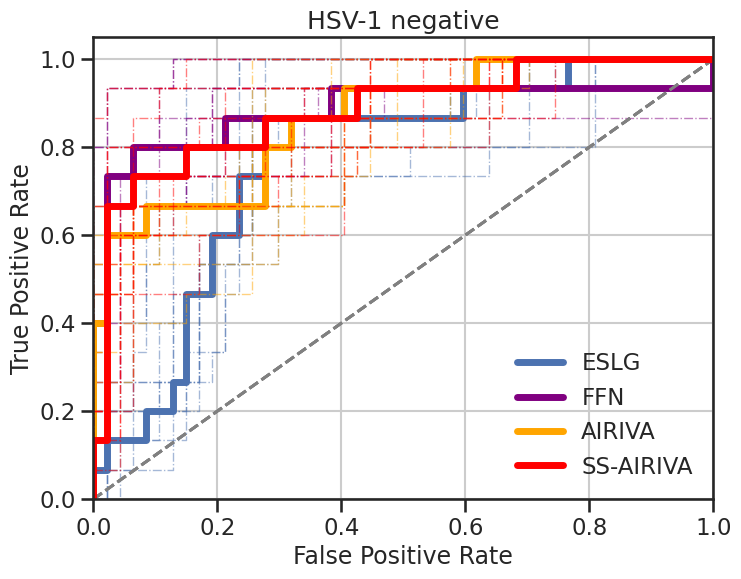}
        \includegraphics[width=0.32\textwidth]{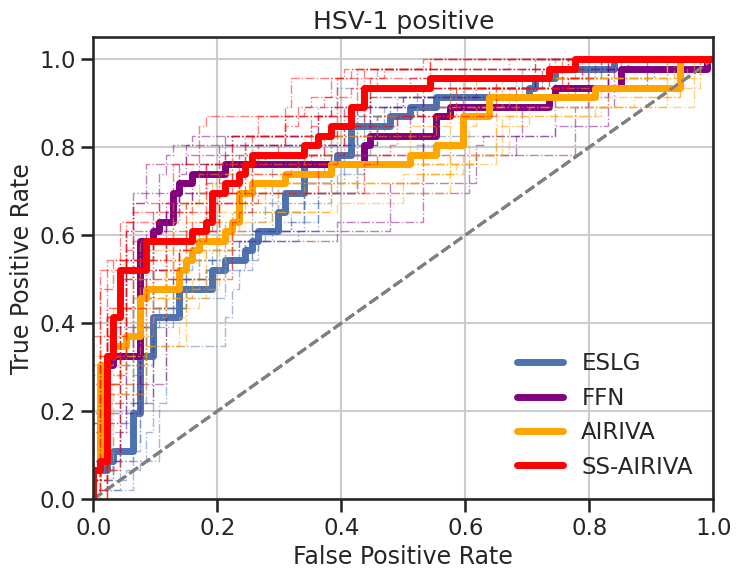}
        \caption*{(b) HSV-2 Prediction Task}\label{fig:hsv2}
    \end{minipage}
    \caption{\textbf{ROC curves for HSV disease prediction models}. Bold lines correspond to the ROC curve on the test set; colored dashed lines correspond to 10 additional bootstrap samples. First row: HSV-1 Prediction Task; Second row: HSV-2 Prediction Task. SS-AIRIVA exhibits the best performance overall and across different subgroups. While ESLG performs poorly at the HSV-1 prediction task for the HSV2-positive subgroup, AIRIVA maintains the same predictive performance overall.}\label{fig:roc_hsv}
\end{figure}

\paragraph{Disease Classification}
Table~\ref{tab:performance_hsv} compares ESLG models trained on HSV-1 and HSV-2 independently against AIRIVA models trained to predict HSV-1 and HSV-2 jointly, either on the labeled repertoires only (AIRIVA), or on all the samples including unlabelled repertoires (SS-AIRIVA). We also compare against FFN, which that predicts both HSV-1 and HSV-2 labels jointly. Figure~\ref{fig:roc_hsv} reports the corresponding ROC curves.
ESLG models show reasonable performance for both HSV-1 and HSV-2 prediction tasks in the overall population (0.62 and 0.75 AUROC respectively), but upon closer inspection, these models do not separate the two labels well. In particular, the HSV-1 model performance drops to AUROC of 0.5 for samples that are HSV-2 positive. In comparison, AIRIVA overall HSV-1 performance is comparable to performance on HSV-2 positives (0.67 AUROC), indicating that the model has learned to disentangle the two labels. In all cases, AIRIVA outperforms the ESLG baseline for HSV-1 both in terms of sensitivity at 98\% specificity and AUROC. 

\paragraph{Learning from Unlabelled Repertoires}
Since AIRIVA is a generative model, we expect that including unlabeled data should improve the model's ability to construct a factorized latent space. Roughly 20\% of our HSV data is missing at least one of the  HSV-1 or HSV-2 labels, see Appendix for details. As expected, including these samples in SS-AIRIVA improves model performance, particularly for the HSV-1 prediction task.

\begin{table}[h!]
    \caption{\textbf{Comparison of HSV disease models}: Sensitivity at 98\% specificity and AUROC, overall and stratified by the performance in the presence of the other subtype, using 100 bootstrap samples. AIRIVA trains with only labeled repertoires. SS-AIRIVA refers to the semi-supervised formulation of AIRIVA, training with additional unlabelled repertoires.}\label{tab:performance_hsv}
    \centering
    \begin{subtable}
    \:{HSV-1 Prediction Task}
    \scriptsize
    \begin{tabular}{ccccccc}
        \toprule
        & \multicolumn{2}{c}{\textbf{Overall} } & \multicolumn{2}{c}{\textbf{HSV-2 negative} } &
      \multicolumn{2}{c}{\textbf{HSV-2 positive} } \\
         \cmidrule(lr){2-3}
        \cmidrule(lr){4-5}
        \cmidrule(lr){6-7}
        HSV-1 Model &  Sensitivity & AUROC  & Sensitivity & AUROC & Sensitivity & AUROC\\
        \midrule
        ESLG & 0.12 $\pm$ 0.10 & 0.62 $\pm$ 0.09 & 0.18 $\pm$ 0.15 & 0.63 $\pm$ 0.12 & 0.14 $\pm$ 0.17 & 0.50 $\pm$ 0.19 \\
        FFN & 0.35 $\pm$ 0.13 & 0.80 $\pm$ 0.05 & 0.45 $\pm$ 0.17 & 0.82 $\pm$ 0.05 & 0.30 $\pm$ 0.20 & 0.72 $\pm$ 0.14 \\
        AIRIVA & 0.30 $\pm$ 0.12 & 0.74 $\pm$ 0.09 & 0.35 $\pm$ 0.20 & 0.74 $\pm$ 0.10 & 0.32 $\pm$ 0.22 & 0.67 $\pm$ 0.16 \\
        SS-AIRIVA & 0.45 $\pm$ 0.18 & 0.79 $\pm$ 0.07 & 0.49 $\pm$ 0.18 & 0.78 $\pm$ 0.09 & 0.53 $\pm$ 0.22 & 0.81 $\pm$ 0.13 \\
        \bottomrule
    \end{tabular}
    \label{tab:performance_hsv1}
    \end{subtable}
    
    \vspace{0.5cm}
    \begin{subtable}
    \:{HSV-2 Prediction Task}
    \scriptsize
    \begin{tabular}{ccccccc}
        \toprule
        & \multicolumn{2}{c}{\textbf{Overall} } & \multicolumn{2}{c}{\textbf{HSV-1 negative} } &
      \multicolumn{2}{c}{\textbf{HSV-1 positive} } \\
         \cmidrule(lr){2-3}
        \cmidrule(lr){4-5}
        \cmidrule(lr){6-7}
        HSV-2 Model &  Sensitivity & AUROC  & Sensitivity & AUROC & Sensitivity & AUROC\\
        \midrule
        ESLG & 0.11 $\pm$ 0.10 & 0.75 $\pm$ 0.07 & 0.16 $\pm$ 0.21 & 0.79 $\pm$ 0.16 & 0.12 $\pm$ 0.15 & 0.75 $\pm$ 0.10 \\
        FFN & 0.38 $\pm$ 0.20 & 0.82 $\pm$ 0.08 & 0.66 $\pm$ 0.29 & 0.88 $\pm$ 0.13 & 0.30 $\pm$ 0.21 & 0.80 $\pm$ 0.09 \\
        AIRIVA & 0.37 $\pm$ 0.18 & 0.78 $\pm$ 0.10 & 0.57 $\pm$ 0.26 & 0.86 $\pm$ 0.12 & 0.32 $\pm$ 0.20 & 0.77 $\pm$ 0.10 \\
        SS-AIRIVA & 0.38 $\pm$ 0.19 & 0.84 $\pm$ 0.06 & 0.53 $\pm$ 0.32 & 0.88 $\pm$ 0.10 & 0.37 $\pm$ 0.25 & 0.84 $\pm$ 0.07 \\
        \bottomrule
    \end{tabular}
    \label{tab:performance_hsv2}
    \end{subtable}
\end{table}



\section{Discussion}\label{sec:limitations}



A trained AIRIVA model is an effective {\bf simulator of TCR repertoires}, providing manipulable ``switches" and ``knobs" that  control  factors of the data-generating process, such as disease and sequencing depth. 
Crucially, these controls act {\em independently} and yield counterfactual repertoires that are {\em consistent}.  For instance, such properties are satisfied when intervening on the disease state only affects disease-associated TCRs and not TCR repertoire depth. Conversely, increasing TCR repertoire depth should affect the prevalence of all TCRs without affecting disease-associated TCRs.  In our case studies, we have shown that we can effectively intervene and generate counterfactual repertoires that add or remove spike/non-spike TCRs, or add HSV-1 TCRs to HSV-2 positive repertoires, without significant changes to predictions of the non-intervened labels. Moreover, we leverage the generated counterfactuals to consistently recover label-specific TCRs when validated with an external dataset of TCR annotations.

The ability of AIRIVA to generate consistent counterfactuals is just one of several benefits of learning a {\bf disentangled} latent representation, each with important potential impact in immunology. With {\bf interpretable factors}, latent-space projections of repertoires could explain diagnosis (predictions), monitor disease factors over time, or pre- and post-treatment.  Our results demonstrate how disentanglement leads to robust predictive performance.  AIRIVA's ability to distinguish natural infection from vaccinated-only repertoires in COVID-19 is one example of real-world clinical impact that interpretable models can provide.  The ability to distinguish between a large panel of, say, infectious diseases  (having multiple and also overlapping sets of TCRs) could have significant clinical impact. From repertoire-level to TCR-level impact, disentangled latent factors can assign TCR-label associations using counterfactuals and CATE computations. Our experiments have validated spike and non-spike binding associations using external data.  By generating CATE on reasonable interventions (see Limitations), AIRIVA can provide impact on both scientific discovery of disease-related TCRs and identification of candidates for further investigation for cellular therapy. 

Finally, while the results of our case-studies demonstrate the value of learning disentangled representations of immunomics data, they are too limited in scope (two labels) to have impact by themselves.  However, because we have validated disentanglement on two-label problem, with overlapping label settings (requiring disentanglement) and because of AIRIVA's improvement with unlabelled repertoires, we believe that AIRIVA can scale to problems with a large number of labels and many varied cohorts.  For example, a natural extension to this work is to expand labels to include a) a large infectious disease panel, b) HLA typing, and c) sequencing batch ids. An AIRIVA model with this set-up could be used to identify HLA-disease associated TCRs (increasing the precision of TCR annotations), as well as batch effects (TCRs that are useful in predicting batch ids).

\paragraph{Limitations}
While we believe AIRIVA is a useful framework for studying TCR repertoire data, we note the following limitations and leave suggested improvements for future work:
\begin{itemize}
    \itemsep0em 
    \item CATE estimation might not be identifiable given more complex immunomics datasets, \emph{i.e.}, HLA-mediated antigen presentation in autoimmune diseases, where HLA status causes exposure to disease-causing antigens. 
    \item Understanding TCR response requires us to model a complex three-body problem: TCR-antigen-HLA. However, for simplicity we avoided modeling HLAs; this could be achieved with AIRIVA by extending the set of labels by HLA status. 
    \item Disentanglement is weakly enforced via two inductive biases: the factorized conditional prior and classifier-guided loss, leading to challenges in training AIRIVA's complex objective function.   Supported by disentanglement literature~\citep{shu2019weakly} and our own experience using counterfactual consistency metrics for model selection, disentanglement could be improved by directly incorporating consistency of counterfactuals into the objective function.
    \item Generating counterfactuals for label combinations that have never been observed in training remains an open problem, which could potentially be addressed by imposing further inductive biases for the decoder to extrapolate to out-of-distribution regions.
    \item Other limitations and weaknesses are: our Poisson likelihood is not the best choice for zero-inflated TCR counts, AIRIVA works best with a limited number of TCRs (ideally should handle much higher numbers robustly); scaling AIRIVA to handle higher dimensional inputs as well as more labels, potentially of lower-quality/with high rates of missingness is an important avenue for future work.
\end{itemize}

\paragraph{Conclusion}
AIRIVA provides a powerful tool for analysing TCR repertoires, for both diagnosis and scientific discovery. 
The inferred latent factors are human-interpretable and can be manipulated to simulate in-silico repertoires and counterfactuals, enabling robust disease diagnosis and identification of disease-specific TCRs.
To our knowledge, this is the first attempt leveraging deep generative models to build a simulator of T-cell Receptor Immune-repertoires and generating in-silico repertoires via conditional and counterfactual generation.
In a similar way that fields like natural language processing and image analysis have been the focus of a large part of the community, we believe that the results of this paper constitute a first step towards the use of advanced machine learning methods in immunomics and will bring awareness about this domain to the ML community.

\bibliography{bib_all}

\appendix
\newpage

\section{Derivations for supervised and semi-supervised AIRIVA}

\subsection{Supervised AIRIVA}

The objective function for supervised AIRIVA\footnote{Note that in~\citep{ilse2020diva} there was an additional domain variable $d$ to account for observed labels against which we desire to be invariant. Since their formulation treats $y$ and $d$ identically, here we have simplified the formulation such that $d$ gets absorbed into a generic vector of labels $y$.} looks as follows:
\begin{eqnarray}
	\text{ELBO}_{l}(x,y) & = & \mathbb{E}_{q(z|x)}[\log p(x|z_y, z_x)] - \beta \, D_{KL}(q(z_x|x) || p(z_x)) - \beta \, D_{KL}(q(z_y|x)||p(z_y|y))  \nonumber \\ 
	\mathcal{L}_{\text{AIRIVA}}	& = & \sum^N_{n=1} \left( \text{ELBO}_{l}(x^{(n)},y^{(n)}) + \alpha \, \mathbb{E}_{q(z^{(n)}|x^{(n)})}[\log q(y^{(n)}|z^{(n)}_y)] \right),
\end{eqnarray}
where $x$ is a high-dimensional vector of TCR counts, $y$ is a vector of repertoire labels, $z = [z_y \,;\, z_x ]$ denote the latent representation of a repertoire, and is composed of predictive latents $z_y$ and residual latents $z_x$, $N$ is the number of available repertoires, and $\text{ELBO}_{l}(x,y)$ refers to the evidence lower bound for a single labelled repertoire.

\paragraph{Derivation of the ELBO for supervised AIRIVA.}

\begin{eqnarray}
	\log p(x,y) & = & \log \int p(x|z) p(z|y) p(y) dz \\
				& = & \log \int p(x|z) p(z|y) \frac{q(z|x)}{q(z|x)}dz \\
			& \geq & \mathbb{E}_{q(z|x)}[ \log p(x|z) + \log p(z|y) - \log q(z|x)] \\
			& \geq & \mathbb{E}_{q(z|x)}[ \log p(x|z_y,z_x) + \log p(z_y|y) + \log p(z_x) \\
			& & \qquad \qquad - \log q(z_x|x) - \log q(z_y|x)] \\
			& \geq & \mathbb{E}_{q(z|x)}[\log p(x|z_y,z_x)] + \mathbb{E}_{q(z|x)}[\log p(z_y|y) - \log q(z_y|x)] \\
			& & \qquad \qquad + \mathbb{E}_{q(z|x)}[\log p(z_x) - \log q(z_x|x)] \\
			& \geq & \mathbb{E}_{q(z|x)}[\log p(x|z_y,z_x)] - D_{KL}(q(z_y|x)||p(z_y|y)) - D_{KL}(q(z_x|x)||p(z_x)),
\end{eqnarray}
where:
\begin{eqnarray}
	\mathbb{E}_{q(z|x)}[\log p(z_x) - \log q(z_x|x)] & = & \mathbb{E}_{q(z_z|x)q(z_x|x)}[\log p(z_y|y) - \log q(z_y|x)] \\
	& = & \mathbb{E}_{(z_y|x)}\left[ \mathbb{E}_{q(z_x|x)} \left[ \log p(z_y|y) - \log q(z_y|x) \right] \right] \\
	& = & \mathbb{E}_{(z_y|x)}\left[ \log p(z_y|y) - \log q(z_y|x) \right].
\end{eqnarray}


\subsection{Semi-supervised AIRIVA}\label{app:ssdiva}

The objective function for semi-supervised AIRIVA as in~\citep{ilse2020diva} looks as follows:
\begin{eqnarray}
	\text{ELBO}_u(x) & = & \mathbb{E}_{q(z|x)}[\log p(x|z_y, z_x)] - \beta \, D_{KL}(q(z_x|x) || p(z_x)) \\
	& & + \, \beta \, \mathbb{E}_{q(z_y|x)q(y|z_y)} \left[ \log p(z_y|y) - \log q(z_y|x) \right] \\
	& & + \, \mathbb{E}_{q(z_y|x)q(y|z_y)} \left[ \log p(y) - \log q(y|z_y) \right]\\
	\mathcal{L}_{\text{SS-AIRIVA}}	& = & \sum^N_{n=1} \text{ELBO}_l(x^{(n)},y^{(n)}) + \sum^M_{m=1} \text{ELBO}_u(x^{(m)},y^{(m)}),
\end{eqnarray}
where $N$ and $M$ refer to the number of labelled and unlabelled repertoires respectively, and $\text{ELBO}_u(x)$ is a lower bound for the marginal likelihood $p(x)$ of an unlabelled repertoire. This expression is not very intuitive and does not account for heterogeneous patterns of missingness, i.e., different labels within $y$ might be missing for different repertoires.

\paragraph{Derivation of semi-supervised AIRIVA}

The probability distribution of an unlabelled repertoire $p(x)$ in AIRIVA can be written as:
\begin{equation}
	p(x) = \int \int \int p(x|z_y,z_x)p(z_y|y)p(z_x)p(y) dz_x d_{z_y} dy.
\end{equation}
We assume the following variational approximation:
\begin{equation}
	q(z_{y},z_{d},z_{x},y) = q(z_{y}|x)q(z_{x}|x)q(y|z_{y}),\\
\end{equation}
where $q(y|z_{y})$ can be parametrized by a categorical distribution $\text{Cat}(y;\pi(z_{y}))$.

\paragraph{ELBO for unlabelled repertoires}

In this section, we re-derive the ELBO for unlabelled repertoires:
	
\begin{eqnarray}
		\log p(x) & = & \,\log \int \int p(x|z_{y},z_{x})p(z_{y}|y)p(z_{x})p(y)d{z} dy\\
		& = & \log \int \int p(x|z_{y},z_{x})p(z_{y}|y)p(z_{x})p(y)\frac{q(z|x)q(y|z_{y})}{q(z|x)q(y|z_{y})}dz dy\\
		& \geq & \mathbb{E}_{q(z|x)q(y|z_{y})} [\log p(x|z_{y},z_{x})+\log p(z_{y}|y) +\log p(z_{x}) + \log p(y) \nonumber \\
		& & \quad\quad\quad\quad - \log q(z|x) - \log q(y|z_{y})] \\
		& \geq & \text{ELBO}_{u}(x)
\end{eqnarray}

\begin{eqnarray}
		\text{ELBO}_{u}(x) & = & \mathbb{E}_{q(z|x)}\left[\log p(x|z_{y},z_{x}) +\log p(z_{x})-\log q(z_{x}|x)\right] \nonumber \\
		& &  + \, \mathbb{E}_{q(z_{y}|x)q(y|z_{y})}\left[\log p(z_{y}|y)-\log q(z_{y}|x)+\log p(y)-\log q(y|z_{y})\right]\\
		 & = & \mathbb{E}_{q(z|x)}\left[\log p(x|z_{y},z_{x})\right] - D_{\text{KL}}(q(z_{x}|x)||p(z_{x})) \nonumber \\
		& & - \, \mathbb{E}_{q(y|z_{y})}\left[D_{\text{KL}}(q(z_{y}|x)||p(z_{y}|y))\right]-\mathbb{E}_{q(z_{y}|x)}\left[D_{\text{KL}}(q(y|z_{y})||p(y))\right]. \label{Eq:elbo_u}
\end{eqnarray}

\paragraph{Unified ELBO}
We can unify the ELBO expression in Eq.~\ref{Eq:elbo_u} for both, labelled and unlabelled cases, even when there are different patterns of missingness per label. Let us assume the extended variational approximation:
	\begin{equation}
	q(y|z_{y})=\left\{ \begin{array}{cc}
		\text{Cat}(y;\pi(z_{y})) & \quad \text{if} \; y \; \text{is not observed}\\
		\delta_{\tilde{y}} & \quad \text{if} \; y=\tilde{y} \; \text{is observed}
	\end{array}\right.
	\end{equation}
	where $\tilde{y}$ is the observed value for the
	random variable $y$ in the case of a labelled repertoires, and $\delta_{\tilde{y}}$ is a Dirac delta at location $\tilde{y}$. The generalized ELBO for semi-supervised learning is: 
	
	\begin{align}
		\text{ELBO}_{\text{SS}}(x,y) & =\, \underbrace{\mathbb{E}_{q(z|x)}\left[\log p(x|z_{y},z_{x})\right]}_{\text{reconstruction loss}}- \beta_x  \, \underbrace{D_{\text{KL}}(q(z_{x}|x)||p(z_{x}))}_{\text{residual KL}} \\
		& - \beta_y \, \left( \underbrace{ \mathbb{E}_{q(y|z_{y})}\left[D_{\text{KL}}(q(z_{y}|x)||p(z_{y}|y))\right]}_{\text{predictive KL}}- \underbrace{\mathbb{E}_{q(z_{y}|x)}\left[D^\dagger_{\text{KL}}(q(y|z_y)||p(y))\right]}_{\text{label KL}} \right),\label{eq:ssl-diva-elbo}
	\end{align}
where $D^{\dagger}_{\text{KL}}(q||p)$ is an extended definition of the KL divergence:
	\begin{equation}
	D^{\dagger}_{\text{KL}}(q||p)=\left\{ \begin{array}{cc}
		0 & \quad \text{if} \; p \; \text{is a Dirac delta} \\
		D_{\text{KL}}(q||p) & \quad \text{otherwise.}\\
		
	\end{array}\right.
\end{equation}

\paragraph{Objective function for semi-supervised AIRIVA}
The semi-supervised AIRIVA objective can then be written as follows:
\begin{equation}
	\mathcal{L_{\text{SS-AIRIVA}}} =\sum_{n=1}^{N+M}\text{ELBO}_{\text{SS}}(s^{(n)})+ \underbrace{\sum_{i\in\mathcal{I}_{y}}\alpha_{y}\mathbb{E}_{q(z^{(i)}_{y}|x^{(i)})}\left[\log q_{\phi}(y^{(i)}|z^{(i)}_{y})\right]}_{\text{predictive loss}},\label{eq:ss-diva}
\end{equation}
where $s_{n}\in\{x_{n},(x_{n},y_{n})\}$
corresponds to the $n_{th}$ repertoire, $N$ and $M$ refer to the number of labelled and unlabelled repertoires respectively, $\text{ELBO}_\text{SSL}$ refers to the unified semi-supervised ELBO from Equation~\ref{eq:ssl-diva-elbo},
and $\mathcal{I}_{y}$ corresponds to the set of indexes for datapoints that have partially observed labels $y$.

\subsection{Comparison with semi-supervised CC-VAE~\citep{joy2020capturing} \label{subsec:Comparison-with-CC-VAE}}\label{app:ssccvae}

\paragraph{Supervised formulation.} DIVA or AIRIVA and CC-VAE rely on the same generative process, but the assumed variational approximations are different. In CC-VAE, the authors	work with the quantity $q(z|x,y)$ and apply Bayes rule to introduce the parametrized networks $q(y|z_{y})$ and $q(z|x)$:
\[
q(z|x,y)=\frac{q(y|z_{y})q(z|x)}{q(y|x)}.
\]
In contrast, the derivation in AIRIVA only considers: $q(y|z_{y})q(z|x)$,
the key difference is the denumerator $q(y|x)$. A drawback from the DIVA derivation is that there is no emergence	of a natural classifier in the supervised case, whereas this arises
naturally in CC-VAE. That means that the DIVA objective function is not a true lower bound anymore, but rather, an approximation to the marginal likelihood of our data.

\paragraph{Semi-supervised formulation.} CC-VAE and AIRIVA exhibit the same formula for the unlabelled case. The objective of both models only differ in the ELBO derivation for the labelled datapoints, in particular:

\begin{equation}
	\text{ELBO}_{\text{AIRIVA}}(x,y)=\mathbb{E}_{q(z_{y}|x)q(z_{x}|x)}\left[\log \frac{p_{\theta}(x|z_{y},z_{x})p(z_{y}|y)p(z_{x})}{q(z_{y}|x)q(z_{x}|x)}\right]
\end{equation}

\begin{align}
	\text{ELBO}&_{\text{CCVAE}}(x,y) = \,\mathbb{E}_{q(z|x,y)}\left[\log \frac{p_{\theta}(x|z_{y},z_{x})p(z_{y}|y)p(z_{x})}{q(z|x,y)}\right] \nonumber\\
	= & \, \mathbb{E}_{q(z_{y}|x)q(z_{x}|x)}\left[\frac{q(y|z_{y})}{q(y|x)}\log \frac{p_{\theta}(x|z_{y},z_{x})p(z_{y}|y)p(z_{x})}{q(y|z_{y})q(z_{y}|x)q(z_{x}|x)}\right]+\log q(y|x)+\text{log} p(y).
\end{align}

 \section{Further Experimental Details}

 \begin{itemize}
     \item To train AIRIVA, we implemented softmax annealing in the variance of the reparameterization trick, which can be interpreted as sampling from a surrogate deterministic version of the posterior $\tilde{q}(z|x)$ with variance increasing with the number of iterations until it matches the variance of the original posterior ${q}(z|x)$.
    \item We restricted the predictive latents associated with each binary label of interest to be scalar variables.
    \item For interpretability, we restrict the regressors of binary labels to be linear, such that any complex feature transformation occurs in the encoder.
    \item We performed a hyperparameter sweeps of 1000 runs on a random grid search over a wide range of parameters, including encoder / decoder architectures, dimensionality of the residual latent space, and objective function weights. The detailed specification for the sweep can be found below:
\end{itemize}

\begin{verbatim}
  alpha_kl_predictive: [1, 10, 100]
  alpha_kl_residual: [1,10,100]
  alpha_predictive: [1,10,100]
  batch_norm: [False, True]
  batch_size: [300, 3000]
  discriminator_lr: [0.001, 0.01, 0.1]
  dropout_prob: [0, 0.4]
  hidden_layers_str: ["", "256", "256,64", "256,128,64"]
  L1_activations: [0, 1, 10]
  L1_embedding_layer: [0, 1, 10]
  learning_rate: [0.0005, 0.001, 0.005]
  num_z_residuals: [10, 50, 100, 300]
  reg_ydecoder: [False, True]
  initialization seed: [5, 6, 7, 8, 1]
  yencoder_hidden_str: ["", "16,4"]
\end{verbatim}

\section{Further Details on the Datasets}

\subsection{COVID-19 Dataset}

\begin{figure}[t]
	\centering
\includegraphics[width=\linewidth]{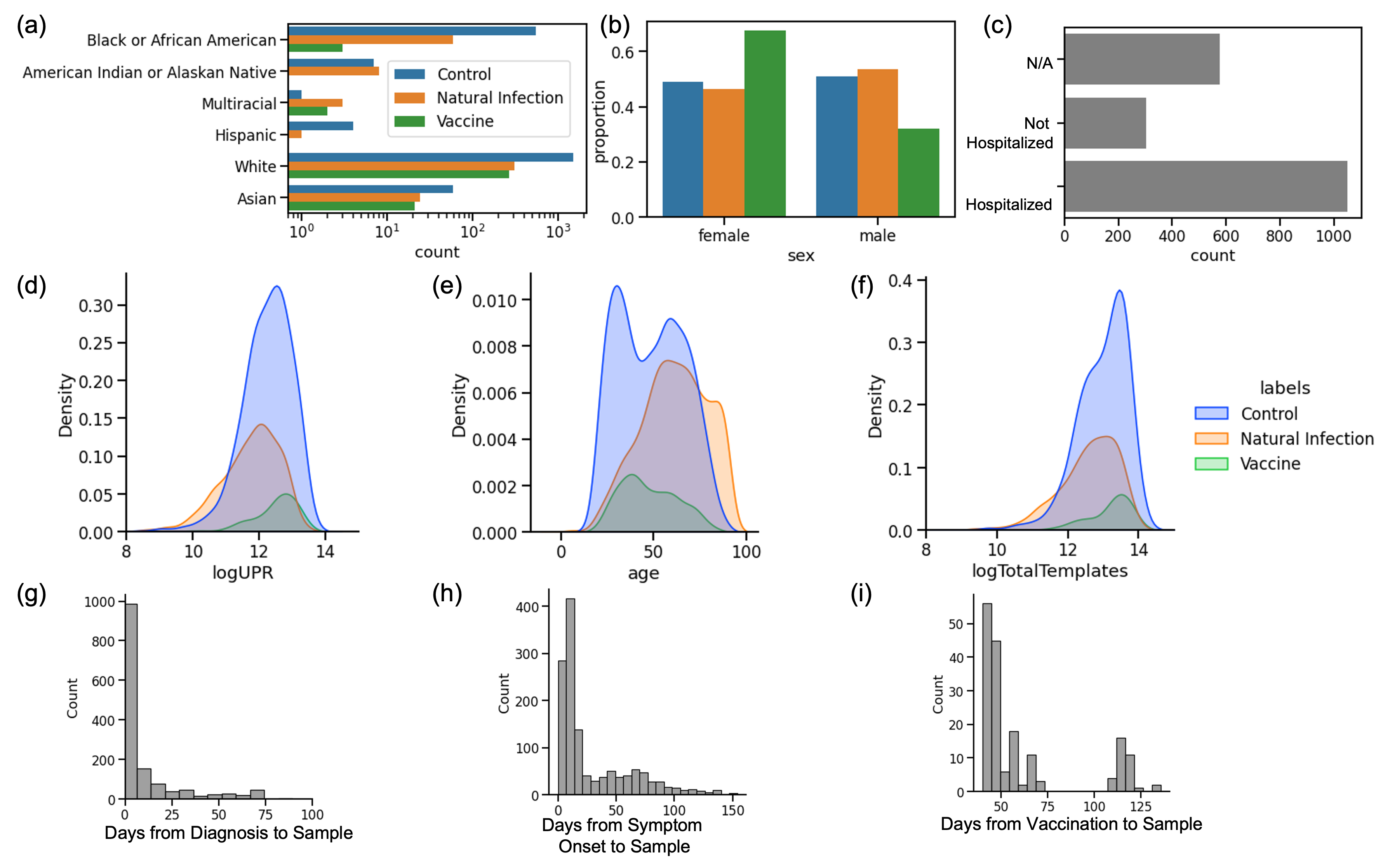}
	\caption{\textbf{COVID-19 Cohort Metadata}  
(a) Race distribution between control, natural infection, and vaccinated cohorts, (b) sex distribution, (c) hospitalization status of natural infection cohort, (d) Log Unique Productive Rearrangement (UPR) distribution, measuring the number of unique T cell clones in a repertoire, (e) age distribution, (f) log total template distribution, measuring the total number of T cell clones in a repertoire,(g) days from diagnosis to blood draw, (h) days from symptom onset to blood draw, (i) days from vaccination to blood draw.}
	\label{fig:meta}
\end{figure}

The COVID-19 diagnostic classifiers are trained on $1,954$ samples from donors with natural SARS-CoV-2 infection, $477$ healthy donors post SARS-CoV-2 vaccination, and $5,198$ healthy controls sampled prior to December 2020. The classifiers are tested on a holdout set of $525$ samples from naturally infected donors, $400$ vaccinated donors, $100$ donors who were naturally infected and later vaccinated, and $4606$ healthy controls. 
The naturally infected donors all tested positive by RT-PCR, with an average of $13$ days from diagnosis to blood draw (Figure \ref{fig:meta}g). The vaccinated donors were sampled following their final scheduled dose---dose 2 for BNT162b2 (Pfizer) and mRNA-1273 (Moderna) and dose 1 for Ad26.COV2.S (Johnson \& Johnson), with an average of $56$ days from vaccination to blood draw (Figure \ref{fig:meta}i).   

The total sequencing depth measured by number of productive T-cell templates is lowest in the natural infection cohort (mean $380$k), with the control and vaccinated cohorts having similar depth (mean $530$k and $540$) (Figure \ref{fig:meta}f). The natural infection and control cohorts are split roughly equally between female and male donors (mean $45\%$ and $50\%$ control), while the vaccinated cohort has more samples from female donors ($61\%$) (Figure \ref{fig:meta}b). Vaccinated and control donors have similar age distributions (mean $47$ years) while donors with natural infection skew slightly older (mean $60$ years) (Figure \ref{fig:meta}e). Of the $43\%$ samples with race metadata, the control cohort is the most diverse, with $42\%$ non-white donors, followed by the natural infection cohort with $36\%$ non-white donors, and the vaccine cohort with $22\%$ non-white donors (Figure \ref{fig:meta}a). 

\subsection{HSV Dataset}
HSV infection labels were created by detecting IgG antibodies against HSV-1 and HSV-2 antigen in the serum of associated TCR repertoires using a multiplexed immunoassay platform (U plex from Meso Scale Discovery). The gG antigen in HSV-1 and gG in HSV-2 were chosen as antigen to detect the presence of HSV-1 and/or HSV-2 antibody responses since they share low sequence homology ($\ll$ 30\%). The gG protein was linked to its respective spot on U-PLEX plates  according to the manufacturer protocol.  Antigen specific antibodies in the serum were detected with SULFO-TAG labeled anti-human IgG antibodies using MSD MESO QuickPlex SQ 120 instrument. Each sample was tested in triplicate.

\begin{table}[t]
    \centering
    \caption{Description of HSV dataset}
    \label{tab:hsv_data}
    \begin{tabular}{cccccc}
    & \textbf{Repertoires} & HSV1+ & HSV2+ & HSV1+ HSV2+ & HSV1- HSV2-  \\
    \midrule
    \textbf{Training} & 1125 & 645 & 295 & 177 & 208  \\
    \textbf{Holdout} & 289 & 169 & 79 & 46 & 47 \\
    \end{tabular}
\end{table}
\begin{table}[t]
    \centering
    \caption{Missing labels in HSV dataset}
    \label{tab:hsv_missings}
    \begin{tabular}{ccccc}
    & \textbf{Repertoires} & HSV-1 missing & HSV-2 missing & Both missing  \\
    \midrule
    \textbf{Training} & 1125 & 193 & 158 & 27 \\
    \textbf{Holdout} & 289 & 50 & 42 & 5 \\
    \end{tabular}
\end{table}

We assigned binary labels to this data set by determining thresholds on the continuous experimental readout by comparing to the background signal level, leaving some samples where we were unable to confidently assign a label. The total number of HSV-1 positives, HSV-2 positives, double positives and unlabeled repertoires are shown in Table~\ref{tab:hsv_data} and Table~\ref{tab:hsv_missings}.

\section{Additional Results for COVID}

\subsection{Spike Prediction Task}

\begin{figure}[th]
    \centering
    \begin{minipage}{0.32\textwidth}
        \centering
        \includegraphics[width=1\textwidth]{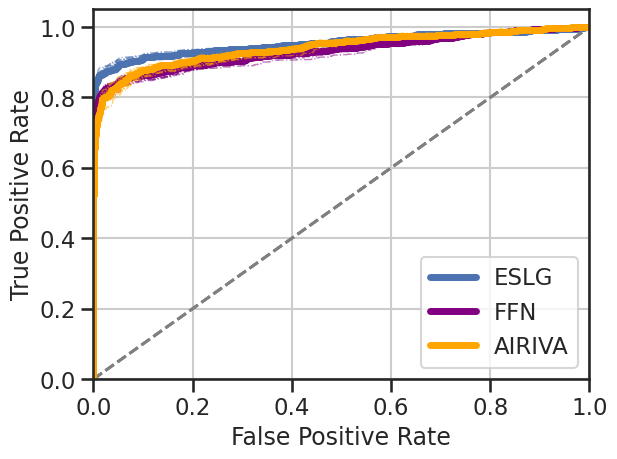}
        \caption*{(a) Overall}\label{fig:roc_spike}
    \end{minipage}
    \begin{minipage}{0.32\textwidth}
        \centering
        \includegraphics[width=1\textwidth]{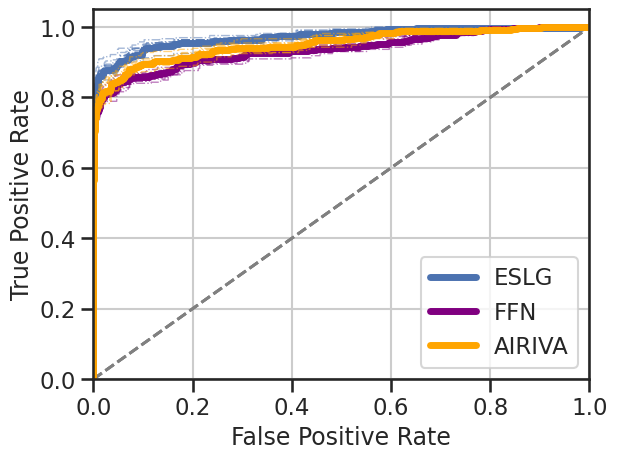}
        \caption*{(b) Healthy}\label{fig:roc_spike_healthy}
    \end{minipage}
    
     \caption{ROC Curves for Spike prediction model.}
     \label{fig:spike_roc}
\end{figure}

Table~\ref{tab:performance_spike} and Figure~\ref{fig:spike_roc} show the performance for spike AIRIVA, ESLG and FFN on the spike prediction task both matching the spike training setup (Overall) and for vaccinated samples versus healthy controls (Healthy). All three models are well matched in performance.

\begin{table}[t]
    \centering
    \caption{Comparison of Spike prediction models: Sensitivity at 98\% specificity and AUCROC) overall and for healthy repertoires.}
    \label{tab:performance_spike}
    \begin{tabular}{ccccc}
        \toprule
        & \multicolumn{2}{c}{\textbf{Overall} } & \multicolumn{2}{c}{\textbf{Healthy} } \\
        \cmidrule(lr){2-3}
        \cmidrule(lr){4-5}
        Model &  Sensitivity & AUROC  & Sensitivity & AUROC\\
        \midrule
        ESLG & 0.87 $\pm$ 0.03 & 0.95 $\pm$ 0.02 & 0.87 $\pm$ 0.05 & 0.97 $\pm$ 0.01 \\
        FFN & 0.82 $\pm$ 0.04 & 0.93 $\pm$ 0.02 & 0.79 $\pm$ 0.06 & 0.93 $\pm$ 0.02 \\
        AIRIVA & 0.79 $\pm$ 0.04 & 0.94 $\pm$ 0.01 & 0.81 $\pm$ 0.06 & 0.95 $\pm$ 0.02 \\
        \bottomrule
    \end{tabular}  
\end{table}

\subsection{Concentrated ROC curves for COVID Disease Model}

\begin{figure}[h!]
   \begin{minipage}{0.32\textwidth}
       \centering
        \includegraphics[width=1\textwidth]{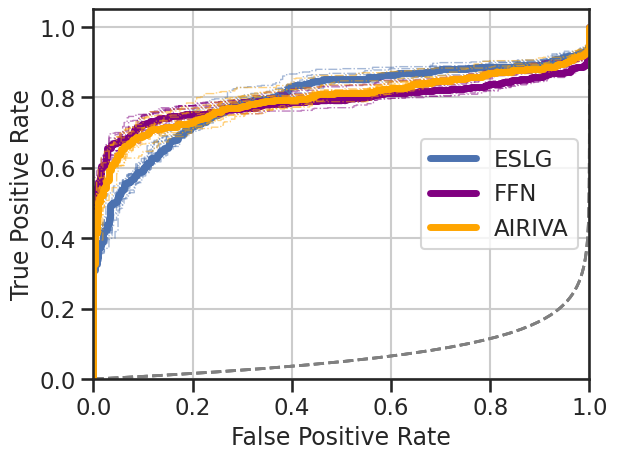}
        \caption*{(a) Overall}\label{fig:croc_covid_all}
    \end{minipage}
    \begin{minipage}{0.32\textwidth}
       \centering
        \includegraphics[width=1\textwidth]{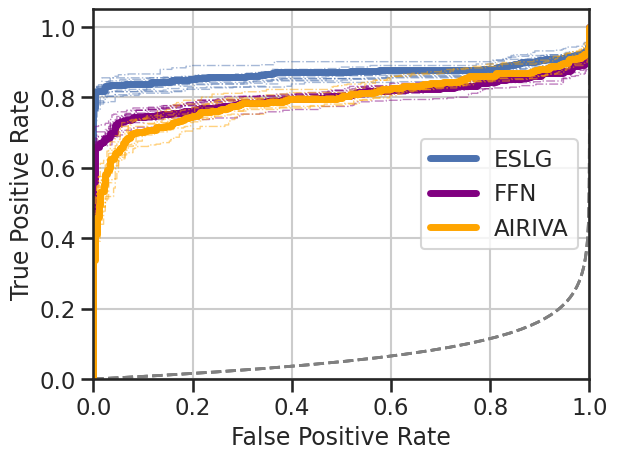}
        \caption*{(b) Unvaccinated}\label{fig:croc_covid_0}
    \end{minipage}
    \begin{minipage}{0.32\textwidth}
       \centering
        \includegraphics[width=1\textwidth]{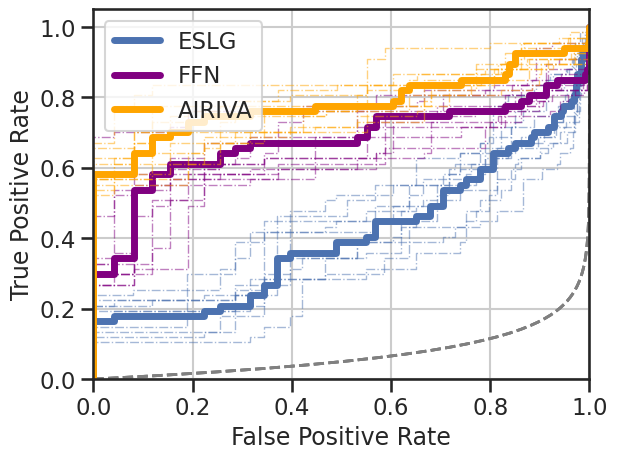}
        \caption*{(c) Vaccinated}\label{fig:croc_covid_1}
   \end{minipage}
    \caption{\textbf{Concentrated ROC Curves~\citep{Swamidass2010} for non-spike disease models.} Bold lines correspond to the CROC curve on the holdout data; colored dashed lines correspond to 10 additional bootstrap samples. ESLG performs poorly for the vaccinated subgroup, while AIRIVA performance remains competitive, with similar predictive performance compared to the overall population.}\label{fig:croc_covid}
\end{figure}

Figure~\ref{fig:croc_covid} shows concentrated ROC curves for the non-spike disease models, matching the same ROC curves reported in the main text.

\begin{figure}[t]
    \centering
    \begin{minipage}[b]{0.52\linewidth}
	\centering
         \includegraphics[width=1\textwidth]{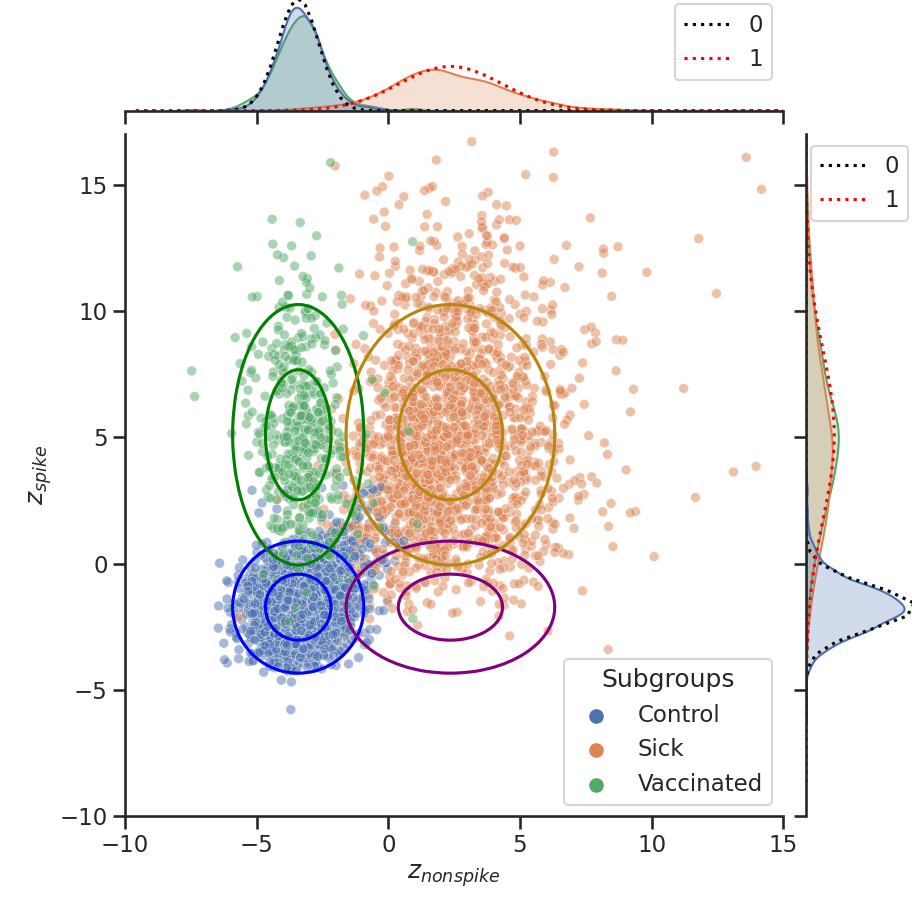}
    \end{minipage}
	\caption{\textbf{Latent space $q(\zv|\xv)$ for COVID-19 in train set}. AIRIVA learns an interpretable latent space that matches biological intuition.
 Each point corresponds to one posterior sample. Predictive priors $p(\zv_y|\yv)$ for each combination of labels are represented as ellipses, each circle denoting one standard deviation. Purple circles correspond to the prior for the \textit{non-spike only} subgroup, which is never observed in reality. On the sides, we show the marginal posteriors as continuous lines, and predictive priors as dashed lines.}
	\label{fig:latent_space_COVID_train}
\end{figure}

\subsection{Latent space for train samples}

Figure~\ref{fig:latent_space_COVID_train} shows the latent space for the COVID-19 training data.

\subsection{Depth Distribution of Conditionally Generated In-silico Repertoires}

\begin{figure}
    \centering
    \includegraphics[width=0.4\linewidth]{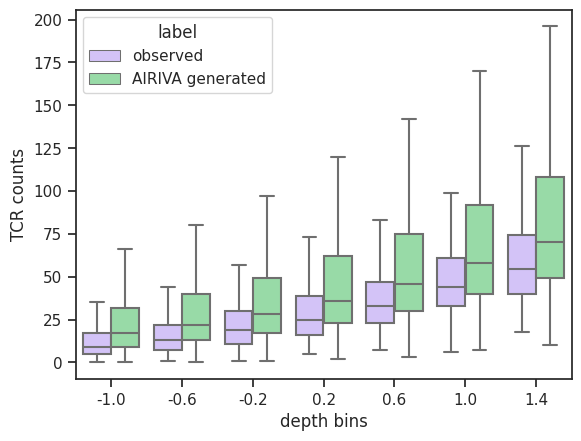}
    \caption{Distribution of TCR counts for observed and in-silico repertoires.}
    \label{fig:effect_interven_depth}
\end{figure}

Figure~\ref{fig:effect_interven_depth} compares the distribution of total sum of TCR counts for conditionally generated in-silico repertoires against the empirical distribution when stratifying by depth. Although the in-silico repertoires tend to exhibit a higher number of outliers due to the assumed Poisson distribution, the marginal distribution of in-silico repertoires follows the same trend than the empirical one, enabling us to control depth in a realistic manner.

\begin{figure}[]
    \begin{minipage}{0.95\textwidth}
       \centering
       \includegraphics[width=\linewidth]{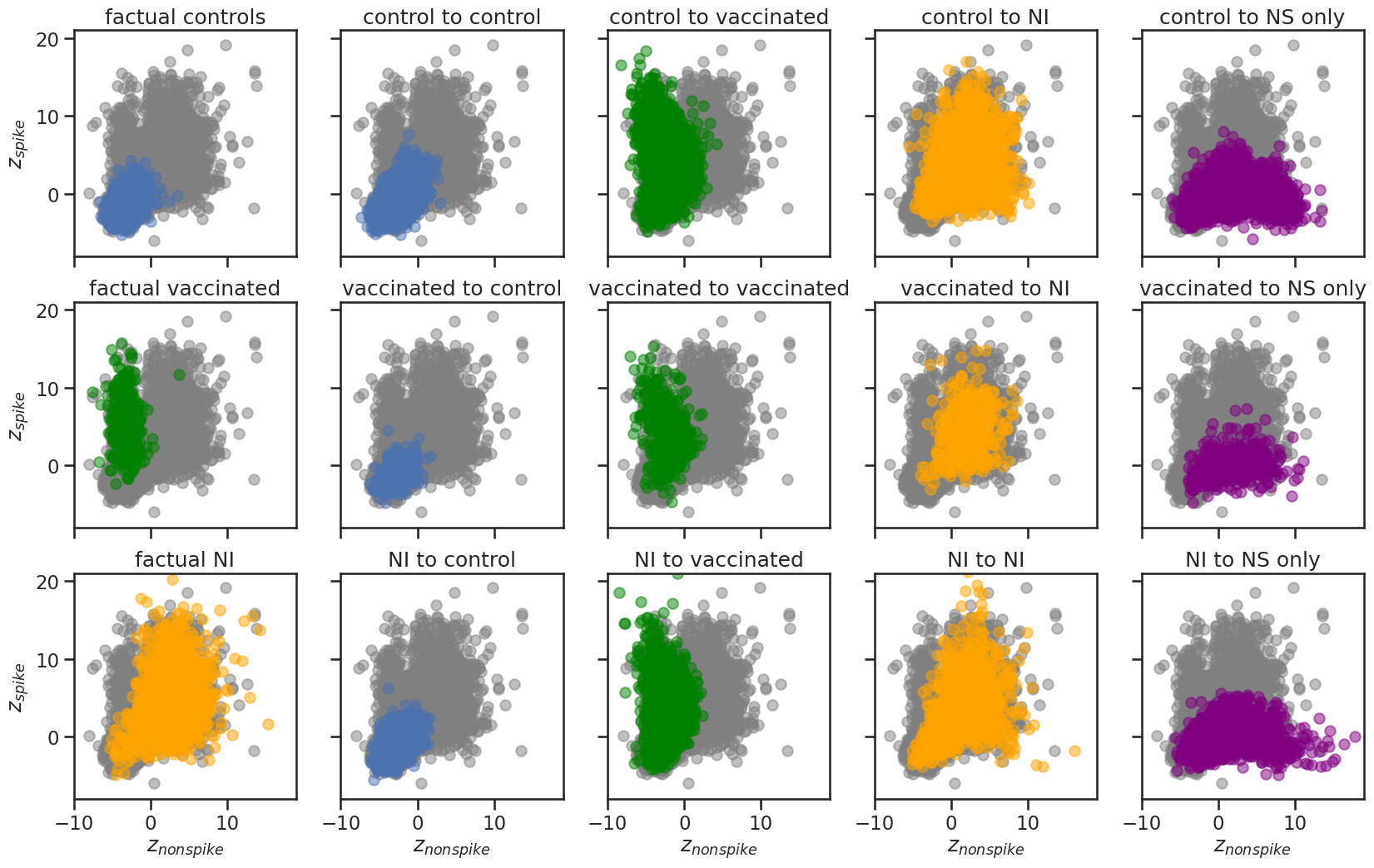}
		\caption*{(a) Posterior samples $\sz \sim q(\zv|\xv)$}
    \end{minipage}
        \centering
     \includegraphics[width=\linewidth]{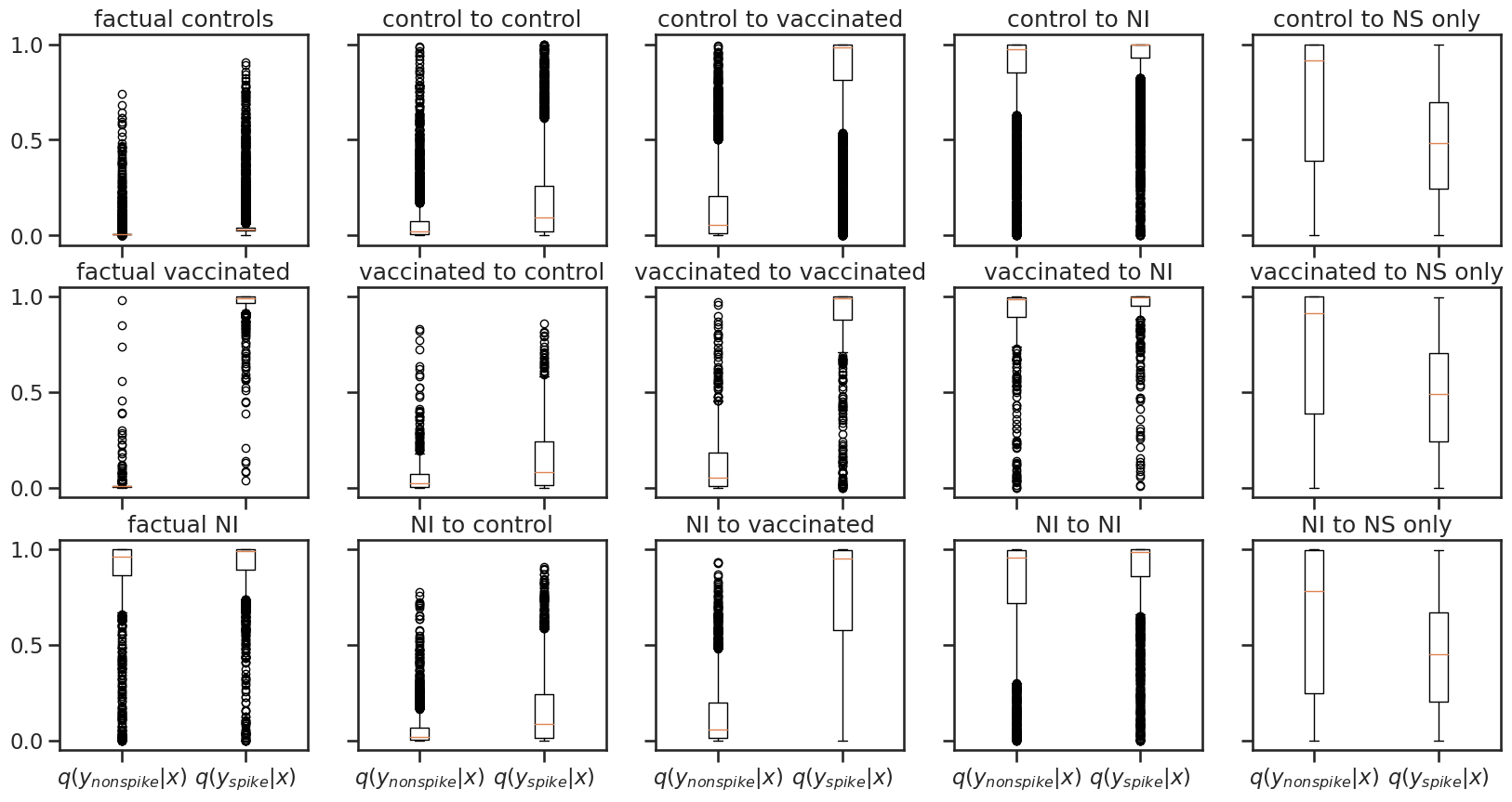}
		\caption*{(b) Posterior predictive $\y \sim q(\yv|\xv)$}
	\caption{\textbf{Counterfactual generation for COVID-19}. (a) Latent representation $\sz \sim q(\zv|\xv)$ for factual data (first column) and counterfactual data (other columns) stratified by subgroups. (b) Posterior predictive distribution $q(\yv|\xv)$ for factual (first column) and counterfactual data (columns 2-5) stratified by subgroups. The labels assignment is: control = [0 0], natural infection [1 0], and vaccinated repertoires [0 1]. The ``nonspike only'' subgroup can never be observed in reality. \
	}
	\label{fig:counterfactuals}
\end{figure}

\subsection{Faithful Counterfactual Generation for COVID-19}

Figure~\ref{fig:counterfactuals} shows posterior and posterior predictive samples for all possible label combinations to generate counterfactuals. This is the complete version of Figure~\ref{fig:counterfactuals_main} in the main text.

\section{Additional Results for HSV}

\begin{figure}[h]
	\centering
	\begin{minipage}{0.44\textwidth}
        \includegraphics[width=\textwidth]{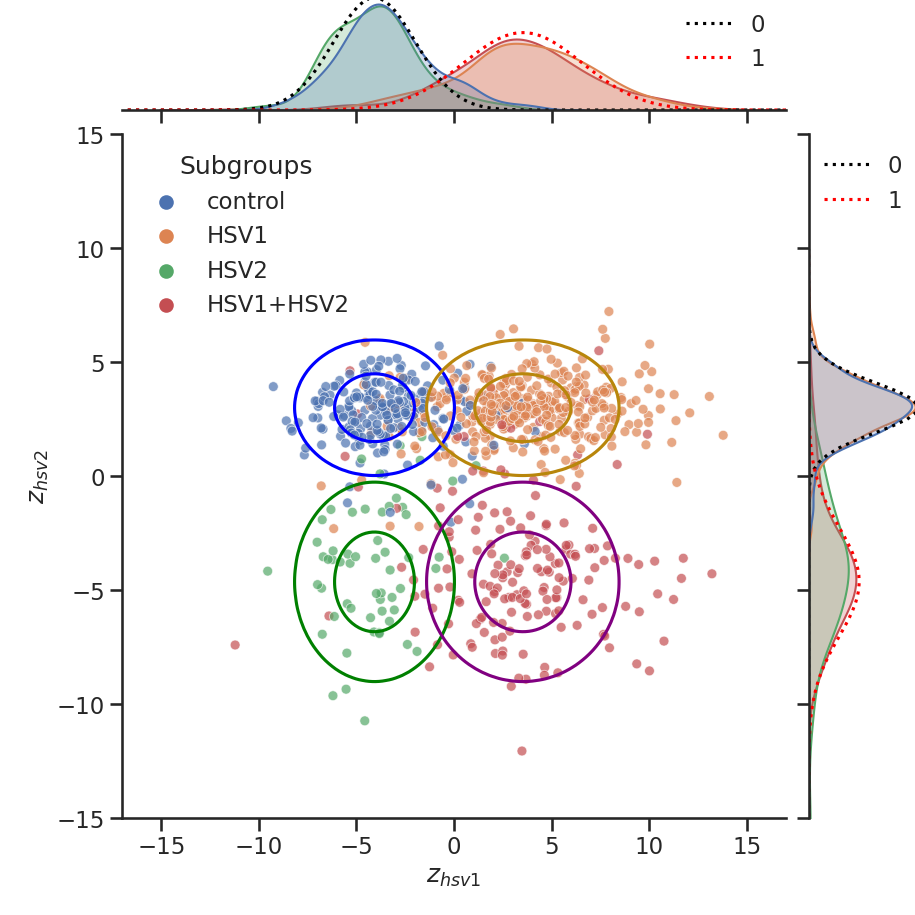}
        \caption*{(a) Train set}
	\end{minipage}
 \begin{minipage}{0.44\textwidth}
        \includegraphics[width=\textwidth]{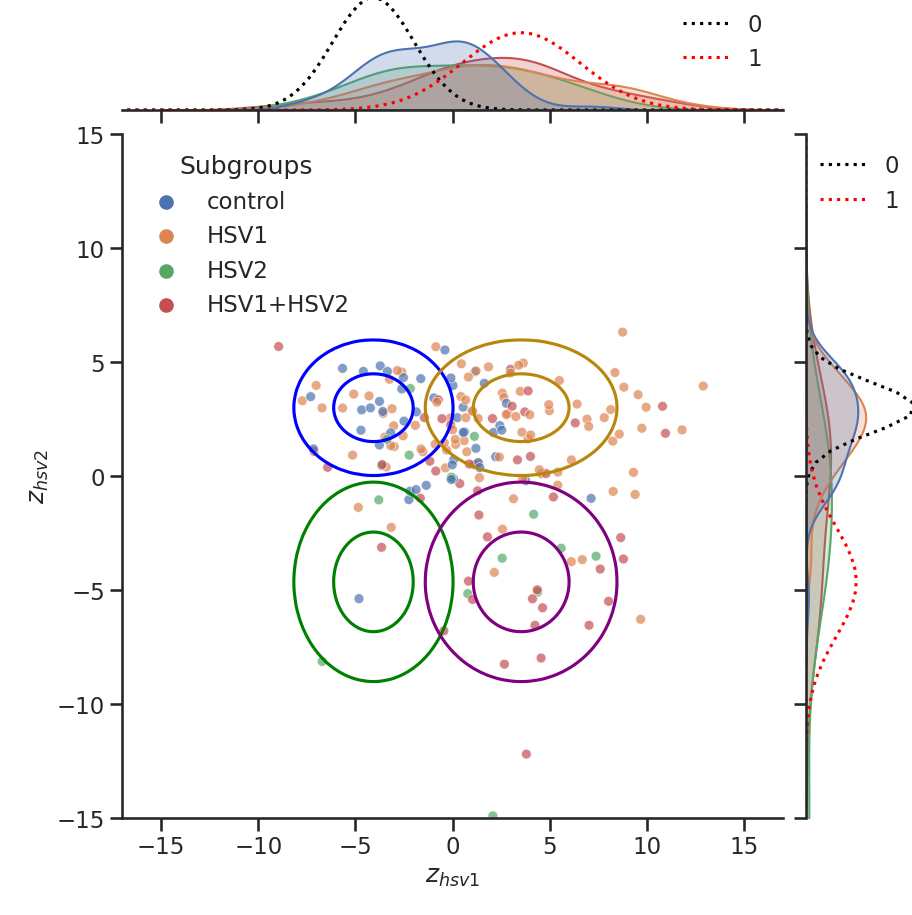}
        \caption*{(b) Holdout set}
	\end{minipage}
	\caption{\textbf{Latent space for HSV}. Each point corresponds to one sample $z$ from the posterior $q(z|x)$ in holdout. On the sides, we show marginal posterior as continuous lines, and predictive priors as dashed lines. Predictive priors $p(z_y|y)$ for each combination of labels $y$ are represented as ellipses, each circle corresponding to one standard deviation.}
	\label{fig:latent_space_hsv12}
\end{figure}

\begin{figure}[]
\centering
    \begin{minipage}{0.80\textwidth}
        \begin{center}
        \includegraphics[width=1\textwidth]{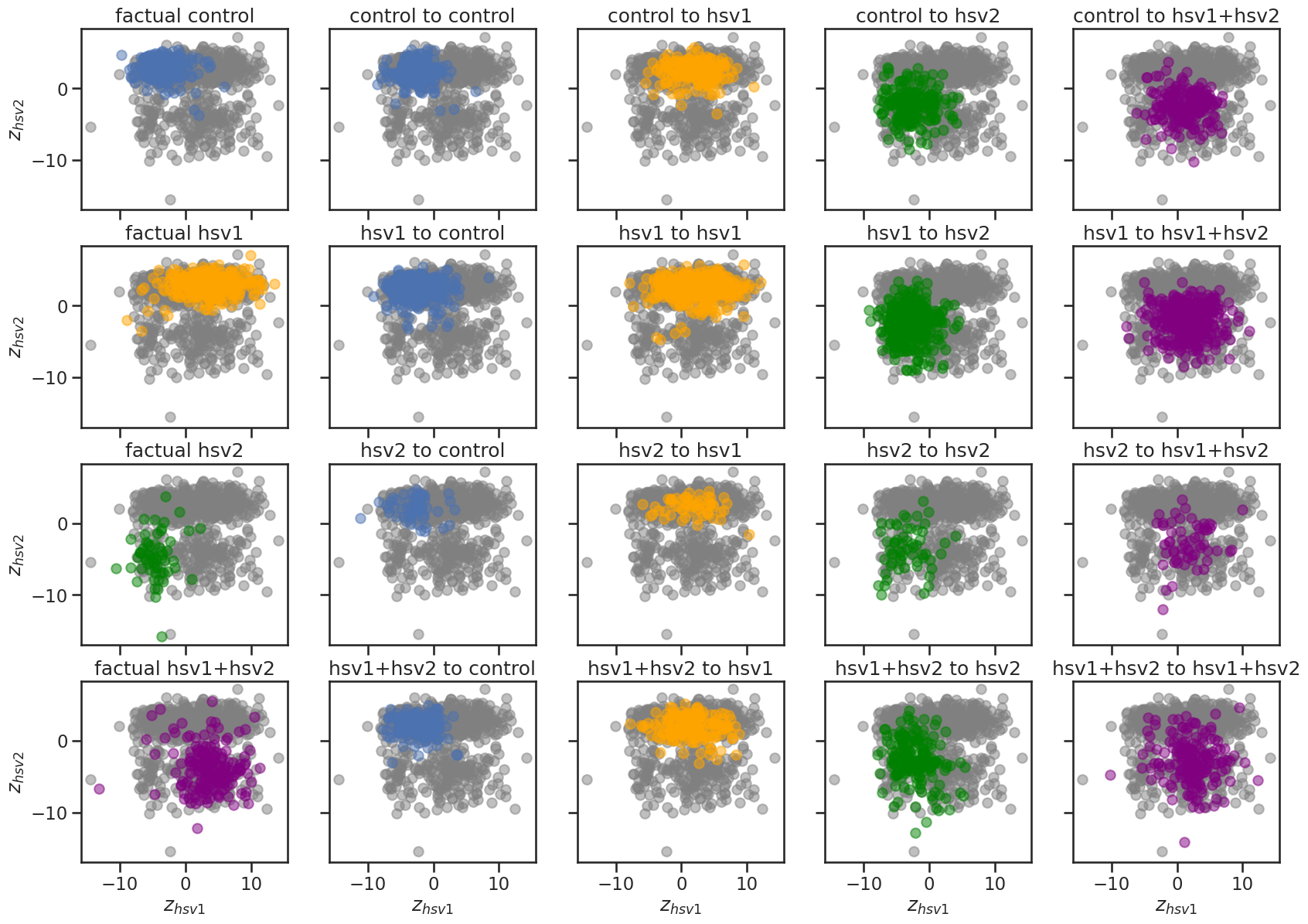}
    \vspace{0.1cm}
		\caption*{(a) Posterior samples $\sz \sim q(\zv|\xv)$}
        \end{center}
    \end{minipage}
    \vspace{0.1cm}
    \begin{minipage}{0.80\textwidth}
        \begin{center}
		\includegraphics[width=1\textwidth]{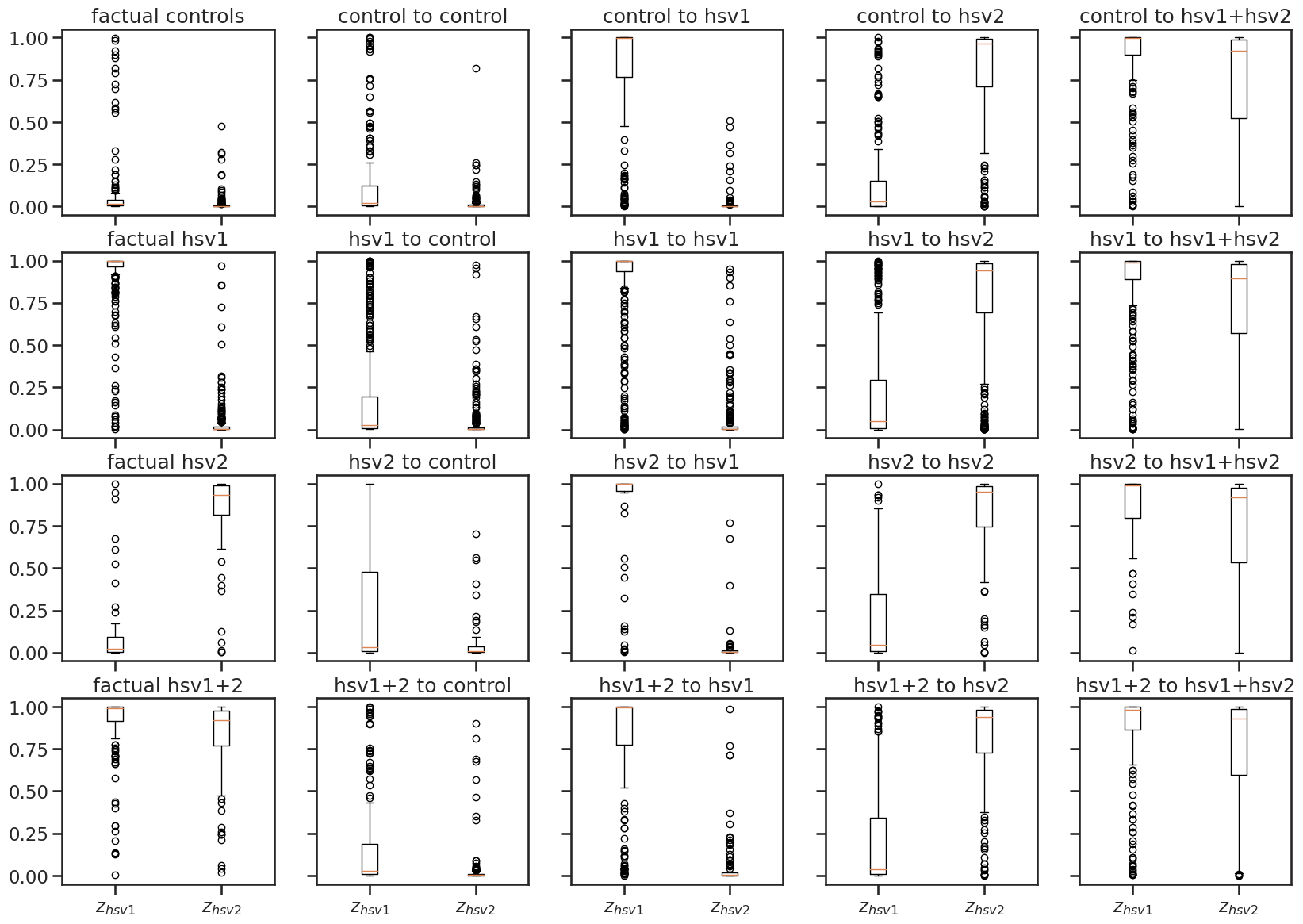}
    \vspace{0.1cm}
		\caption*{(b) Posterior predictive $\y \sim q(\yv|\xv)$}
        \end{center}
    \end{minipage}
	\caption{\textbf{Counterfactual generation for HSV}. (left) Latent representation $\sz \sim q(\zv|\xv)$ for factual data (first column) and counterfactual data (other columns) stratified by subgroups. (right) Posterior predictive distribution $q(\yv|\xv)$ for factual (first column) and counterfactual data (columns 2-5) stratified by subgroups. Label assignments are: control = [0 0], HSV1 [1 0], HSV2 [0 1] and HSV1+HSV2 [0 1].
	}
	\label{fig:counterfactuals_hsv}
\end{figure}

Figure~\ref{fig:latent_space_hsv12} shows the latent space representation for AIRIVA trained on HSV. We see changes in HSV-1/HSV2 status of repertoires along the corresponding predictive latent axes. Figure~\ref{fig:counterfactuals_hsv} shows counterfactuals for each HSV subtype. Similarly to the COVID-19 case study, AIRIVA is able to generate counterfactuals whose latent representation is consistent with the latent representation of the observed data. Note that unlike COVID-19, all subgroups are observed.

\end{document}